\PassOptionsToPackage{protrusion, expansion}{microtype} %% for better typography
\PassOptionsToPackage{dvipsnames}{xcolor} %% for better typography
\documentclass[conference, nofonttune]{IEEEtran}\def\UseIEEETemplate{1}  %% my macro
\IEEEoverridecommandlockouts%% intended to improve IEEE author block

\usepackage{graphicx}
\usepackage{xcolor}
\definecolor{B}    {HTML}{2b66d3}   %% blue, light
\definecolor{B2}   {HTML}{003399}   %% blue, dark 
\definecolor{Bv}   {HTML}{0000EB}   %% blue, vibrant
\definecolor{R}    {HTML}{c9171e}
\definecolor{R2}   {HTML}{d7003a}
\definecolor{INK}  {HTML}{595857}
\definecolor{Y}    {HTML}{f1c40f}
\definecolor{G}    {HTML}{009a00}
\definecolor{GRAY} {HTML}{808080}
\definecolor{MAUVE}{HTML}{9400D1}

\newcommand{\cusz}{{cuSZ}}       %% change to scshape when camera-ready
\newcommand{\thiswork}{\mbox{{cuSZ-I}}}  %% hyphen non-breakable
\newcommand{\ginterp}{\mbox{\textit{G-Interp}}} %% hyphen non-breakable

\usepackage{array}      %% https://tex.stackexchange.com/q/338009
\usepackage{booktabs}
\usepackage{colortbl}
\usepackage{adjustbox}
\usepackage{multirow}
\usepackage{tabu}       %% friendly when coloring

\usepackage{url, hyperref}                  %% hyperlink
\hypersetup{
    colorlinks, linkcolor=RoyalBlue, citecolor=RoyalBlue, urlcolor=RoyalBlue,
    allcolors=RoyalBlue,
    pdfborder={0 0 0}}

\usepackage{enumitem}                       %% enhanced items
\usepackage{lipsum}                         %% pseudo texts
\usepackage[plain]{algorithm}       		% algo%rithm
\floatstyle{plaintop}
\usepackage{algpseudocode}
\algrenewcommand{\alglinenumber}[1]{{\scriptsize\bfseries\ttfamily\color{R}#1}}
\floatstyle{plaintop}
\restylefloat{algorithm}

\usepackage{xpatch}% specifically for fixing indentation, thanks to https://tex.stackexchange.com/questions/294840/indent-algorithmic-outside-of-a-block
\makeatletter
\xpatchcmd{\algorithmic}{\ALG@tlm\z@}{\ALG@tlm\z@\leftmargin 10pt}{}{}
\makeatother

\usepackage{listings}						%% code listing
\usepackage{setspace}                       % spacing

\usepackage{soul}%% \hl{} and \st{}
\ifx\UseACMTemplate\undefined
\else
\fi

\ifx\UseIEEETemplate\undefined
\else
\fi

\colorlet{HLCOLOR}{B}
\colorlet{TableAltColor}{gray!20}

\usepackage{tikz}
\usetikzlibrary{positioning}        %% must-have
\usetikzlibrary{shapes.multipart}   %% multiline node, should use as default
\usetikzlibrary{shapes.geometric}   %% diamond etc
\usetikzlibrary{arrows}
\usetikzlibrary{arrows.meta}        %% e.g., {[left]Bar}-{[right]Bar}
\usetikzlibrary{backgrounds}
\usetikzlibrary{patterns}

\usepackage{pgfplots}
\usepackage{pgfplotstable}
\usepgfplotslibrary{groupplots}

\usepackage{amsmath}    %% \text in mathmode
\usepackage{amsfonts}
\usepackage{amsthm}
\usepackage{mathtools}
\usepackage{bm}

\usepackage[labelfont=bf,textfont=bf]{caption}
\usepackage{subcaption}
\usepackage[normalem]{ulem}
\usepackage{stackengine}
\usepackage{rotating}	%% dependency of turn

\usepackage[scale=.92]{tgheros}

\hypersetup{
	pdftitle={TurboFFT: A High-Performance Fast Fourier Transform with Fault Tolerance on GPU},
	pdfsubject={TurboFFT, submission},
	pdfkeywords={GPU; HPC}
}

\usepackage[style=ieee, citestyle=numeric, sortcites=true, backend=bibtex, maxbibnames=99]{biblatex}
\usepackage{enumitem}
\usepackage{algorithm}
\usepackage{algpseudocode}
\usepackage{multicol}
\usepackage{multirow}
\usepackage{caption}
\usepackage{subcaption}
\usepackage{subcaption}
\usepackage{booktabs} % For nice tables
\usepackage{siunitx} % To align table numbers by unit
\usepackage{etoolbox}
\usepackage{soul}
\usepackage{cancel}
\usepackage[normalem]{ulem}

\usepackage[font=small,skip=0pt]{caption}

\usepackage{amsmath,algorithm,tabularx}
\usepackage{algpseudocode}

\makeatletter
\newcommand{\multiline}[1]{%
  \begin{tabularx}{\dimexpr\linewidth-\ALG@thistlm}[t]{@{}X@{}}
    #1
  \end{tabularx}
}

\makeatother
\AtBeginDocument{%
  \providecommand\BibTeX{{%
    \normalfont B\kern-0.5em{\scshape i\kern-0.25em b}\kern-0.8em\TeX}}}

\newcommand{\EDIT}[2]{%
    \bgroup%
    \colorbox{B}{\color{white}#1:}\color{B} #2
    \egroup%
}

\pgfplotsset{compat=1.15} 

\begin{document}
\title{TurboFFT: A High-Performance Fast Fourier Transform with Fault Tolerance on GPU\vspace{-0mm}}

%% change when camera-ready
% \author{(author list placeholder: double-blinded)}
% \author{}
% \newcommand{\CoMark}{$^{(\star)}$}
% \newcommand{\IuMark}{$^\circ$}
\newcommand{\UcrMark}{$^\ddagger$}
\newcommand{\AnlMark}{$^\dagger$}

\author{%
    Shixun Wu\UcrMark,
    Yujia Zhai\UcrMark,
    Jinyang Liu\UcrMark,
    Jiajun Huang\UcrMark,
    Zizhe Jian\UcrMark,
    Huangliang Dai\UcrMark, \\
    Sheng Di\AnlMark,
    Zizhong Chen\UcrMark,
    Franck Cappello\AnlMark
    \\[1ex]
    % \IEEEauthorblockA{\CoMark Co-first Authors}
    \IEEEauthorblockA{\UcrMark University of California, Riverside, Riverside, CA, USA}
    % \IEEEauthorblockA{\IuMark Indiana University, Bloomington, IN, USA}
    \IEEEauthorblockA{\AnlMark Argonne National Laboratory, Lemont, IL, USA}
    % \IEEEauthorblockA{\UiowaMark The University of Iowa, Iowa City, IA, USA}
    % \IEEEauthorblockA{\FsuMark Florida State University, Tallahassee, FL, USA}
    swu264, yzhai015, jliu447, 
    jhuan380, zjian106, hdai022@ucr.edu, \\ sdi1@anl.gov, chen@cs.ucr.edu, cappello@mcs.anl.gov
}
%% comment before submission; only for counting pages
\thispagestyle{plain}\pagestyle{plain}

\maketitle
\begin{abstract}
The Fast Fourier Transform (FFT), as a core computation in a wide range of scientific applications, is increasingly threatened by reliability issues. In this paper, we introduce TurboFFT, a high-performance FFT implementation equipped with a two-sided checksum scheme that detects and corrects silent data corruptions at computing units efficiently. The proposed two-sided checksum addresses the error propagation issue by encoding a batch of input signals with different linear combinations, which not only allows fast batched error detection but also enables error correction on-the-fly instead of recomputing. We explore two-sided checksum designs at the kernel, thread, and threadblock levels, and provide a baseline FFT implementation competitive to the state-of-the-art, closed-source cuFFT. We demonstrate a kernel fusion strategy to mitigate and overlap the computation/memory overhead introduced by fault tolerance with underlying FFT computation. We present a template-based code generation strategy to reduce development costs and support a wide range of input sizes and data types. Experimental results on an NVIDIA A100 server GPU and a Tesla Turing T4 GPU demonstrate TurboFFT offers a competitive or superior performance compared to the closed-source library cuFFT. TurboFFT only incurs a minimum overhead (7\% to 15\% on average) compared to cuFFT, even under hundreds of error injections per minute for both single and double precision. TurboFFT achieves a 23\% improvement compared to existing fault tolerance FFT schemes.
\end{abstract}

% A single error in FFT leads to an exponential increase in the total number of errors. Due to the error propagation, existing fault tolerance schemes not only necessitate a checksum computation for each signal and a time-redundant recompute for error correction but also lack architecture-aware optimization on GPU. 
%%
%% Keywords. The author(s) should pick words that accurately describe
%% the work being presented. Separate the keywords with commas.
% \keywords{FFT, GPU, Performance Optimization, Reliability, Resilience}

%% A "teaser" image appears between the author and affiliation
%% information and the body of the document, and typically spans the
%% page.

%%
%% This command processes the author and affiliation and title
%% information and builds the first part of the formatted document.

% \settopmatter{printfolios=true}

% \ccsdesc[500]{Computing methodologies~Massively parallel algorithms}

% \maketitle
\section{Introduction}
The Fast Fourier Transform (FFT), as a core computation in a wide range of applications, is increasingly threatened by silent data corruption. FFT is employed extensively in science and engineering, such as exascale projects like LAMMPS \cite{thompson2022lammps}, quantum simulations like QMCPACK \cite{kim2018qmcpack}, molecular dynamics like HACC \cite{habib2016hacc}. 
A significant portion, e.g. 70\%, of processing time in scientific applications is consumed by FFT, showcased by a space telescope project \cite{stockman1999ngst, murphy1998ngst}. However, those applications are increasingly vulnerable to transient faults caused by high circuit density,  low near-threshold voltage, and low near-threshold voltage \cite{lutz1993analyzing, nicolaidis1999time,laprie1985dependable}. Oliveira et al. \cite{oliveira2017experimental} demonstrated an exascale
system with 190,000 cutting-edge Xeon Phi processors
that still suffer from daily transient errors under ECC protection. Recognizing the importance of this issue, the U.S. Department of Energy has named reliability as a major challenge for exascale computing \cite{lucas2014doe}.

% The significant influence of transient faults is well-documented with numerous instances of server crashes and entire clusters unusable \cite{may1979alpha,baumann2002soft,geist2016supercomputing,hochschild2021cores,dixit2021silent}.  

% Microprocessors are more vulnerable to transient faults due to performance-enhancing technologies like increased circuit density, lower near-threshold voltage, and shrinking transistor size \cite{laprie1985dependable,lutz1993analyzing,nicolaidis1999time}. Although the probability of data corruption occurring in a single computing unit is low, a large-scale cluster still suffers from daily data corruption \cite{oliveira2017experimental}. Under the silent data corruption, computational programs unknowingly transition into undefined states, leading to significant computational errors, invalidating the complete computation results, and even causing the crash of the program or the entire computing cluster \cite{baumann2002soft,geist2016supercomputing}. Recognizing the importance of this issue, the U.S. Department of Energy has named reliability as a major challenge for exascale computing \cite{lucas2014doe}.

Since Intel Corporation first reported a transient error and the consequent soft data corruption in 1978  \cite{may1979alpha}, both the academic and industry sectors have recognized the significant impact of transient faults. In 2000, Sun Microsystems identified cosmic ray strikes on unprotected caches as the cause of server crashes that led to outages at major customer sites, including America Online and eBay \cite{baumann2002soft}. In 2003, Virginia Tech broke down and sold online its newly-built Big Mac cluster of 1100 Apple Power Mac G5 computers because it lacked ECC protection, leading to unusability due to cosmic ray-induced crashes \cite{geist2016supercomputing}. Despite ECC protection, transient faults remain a threat to system reliability. For instance, Oliveira et al. simulated an exascale system with 190,000 cutting-edge Xeon Phi processors, finding it still vulnerable to daily transient errors under ECC \cite{oliveira2017experimental}. Such faults have not only been a concern in simulations; Google has encountered transient faults in its real-world production environment, resulting in incorrect outputs \cite{hochschild2021cores}. In response to the persistent challenge posed by transient faults on large-scale infrastructure services, Meta initiated an internal investigation in 2018 to explore solutions \cite{dixit2021silent}.

% The Fast Fourier Transform (FFT) is ranked among the top 10 algorithms of the 20th century \cite{dongarra2000guest} and is crucial for a wide range of applications from electronics and molecular simulation to astronomical observation and interstellar communication \cite{stockman1999ngst,murphy1998ngst,ayala2021interim}. FFT occupied more than $70\%$ of the computation in the Next Generation Space Telescope's phase retrieval algorithm used in NASA’s Remote Exploration and Experimentation project \cite{stockman1999ngst,murphy1998ngst}.

Transient faults can result in fail-stop errors, causing crashes, or fail-continue errors, leading to incorrect results. Checkpoint/restart mechanisms \cite{phillips2005scalable, NEURIPS2019_9015, tao2018improving, tensorflow2015-whitepaper} or algorithmic methods \cite{hakkarinen2014fail, chen2008scalable, chen2008extending} can often mitigate fail-stop errors, whereas fail-continue errors pose a greater risk by silently corrupting application states and yielding incorrect result \cite{mitra2014resilience, cher2014understanding, dongarra2011international, calhoun2017towards, snir2014addressing}. These errors can be especially challenging in safety-critical scenarios \cite{li2017understanding}. In this paper, we concentrate on fail-continue errors occurring in computing logic units, such as incorrect outcomes, and assume that memory and fail-stop errors are addressed through error-correcting codes and checkpoint/restart. We describe these types of errors as soft errors.

To protect FFT against soft errors, a variety of fault tolerance methods have been proposed \cite{may1979alpha, baumann2002soft,geist2016supercomputing}. Antola et al. developed a time-redundant approach \cite{antola1992fault}. Choi and Malek introduced an FFT fault tolerance scheme based on recomputation through an alternate path \cite{choi1988fault}. Jou and Abraham suggested an algorithm-based fault tolerance (ABFT) method for FFT networks that provides full fault coverage and throughput with a hardware overhead of $O(\frac{2}{\log_2N})$ \cite{jou1988fault}. Later, Tao and Hartmann proposed a novel encoding scheme for FFT networks, enhancing fault coverage by adding 5\% more hardware \cite{tao1993novel}. Wang and Jha introduced a new concurrent error detection (CED) method that achieves improved results with lesser hardware redundancy \cite{wang1994algorithm}. Adopting a similar logic, Oh et al. presented a CED scheme employing a different checksum for increased fault coverage \cite{oh1993algorithm}. Additionally, advancements have been made in parallel systems and GPUs. Banerjee et al. proposed a fault-tolerant design for hypercube multiprocessors \cite{banerjee1990algorithm}. Pilla et al. outlined specific software-based hardening strategies to lower failure rates \cite{pilla2014software}. Fu and Yang implemented a fault-tolerant parallel FFT using MPI \cite{fu2009fault}. Recently, Xin et al. integrated a fault tolerance scheme into FFTW \cite{liang2017correcting}.

However, existing fault tolerance schemes for FFT suffer from 1) per-signal checksum error detection, 2) time-redundant recompute-based error correction, and 3) the lack of architecture-aware optimization. Firstly, error detection employs a non-trivial encoding vector to the checksum input/output signal, resulting in additional overhead in storing the encoding vector and computing the checksum. Secondly, error correction requires a time-redundant recompute because of the exponential propagation of corruption as illustrated in Figure 1, namely one corrupted data infects two output data during one butterfly operation. Finally, existing works focus on the fault tolerance side, while neglecting the architectural-aware optimizations for FFT computation, resulting in resource unbalance and inefficiency. To address these issues, we propose TurboFFT,  a high-performance FFT implementation equipped with an algorithm-based fault tolerance scheme that detects and corrects silent data corruptions at computing units on-the-fly. More specifically, our contributions include the following:

% Even one of the state-of-the-art open-source libraries, VkFFT, still incurs a maximum $30\%$ overhead compared to the closed-source state-of-the-art library, cuFFT.

\begin{figure}[t!]
    \centering
    \includegraphics[width=0.7\linewidth]{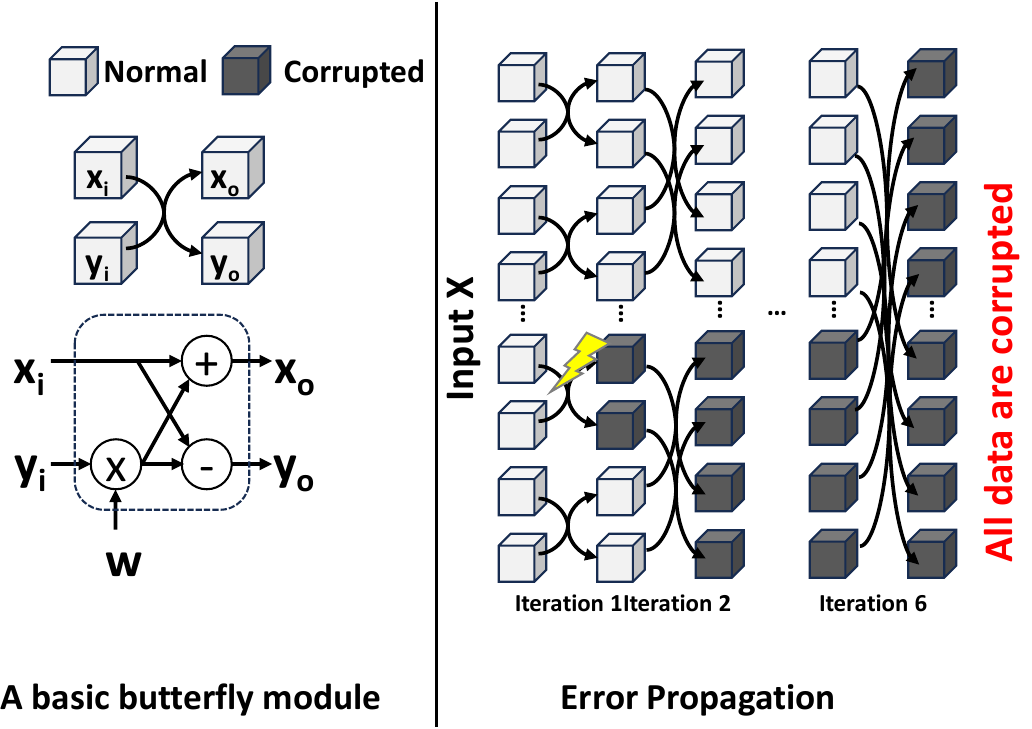}
    \caption{Error Propagation: One corrupted
 propagates to two outputs. The number of data corruptions increases exponentially.}
    \label{fig:err_prop}
    \vspace{-1mm}
\end{figure}

% Existing fault tolerance mechanisms for FFT could not be effectively applied to the hierarchical FFT algorithm on GPUs due to two major challenges: \textit{the memory-bound nature of FFT} and \textit{the exponential growth of error propagation}. 
% FFT is a fundamental algorithm integral to a wide range of applications in scientific computing, typically executed on large-scale computing facilities over extended periods. As a result, the need for resilience against FFT soft errors on GPUs is critical. However, prior work mainly focused studied on CPUs \cite{chen2008algorithm, gunnels2001fault, wu2013line} while the fault tolerant FFT on GPU is still under-studied. Existing works, which have demonstrated the feasibility to deploy ABFT, either lack the architectural-aware designs for GPUs or are lack of optimizations to mitigate the memory cost introduced by ABFT \cite{ding2011matrix}. To address these issues, we first build TurboFFT, a high-performance FFT kernel from scratch, which serves as the foundation for our novel and efficient approach that fuses the memory operations of ABFT with the original FFT memory footprint through the implementation of custom fault-tolerant GPU kernels. We have also designed a template-based code generation scheme that eases development costs to minimize the overhead of fault tolerance functionality in TurboFFT for different input sizes. More specifically, our contributions include the following:

\begin{itemize}[leftmargin=*]

\item We address the error propagation issue by proposing a novel two-sided checksum scheme for FFT. Not only does our proposed algorithm overcome the problem of error propagation but it also enables on-the-fly fault correction, eliminating the need for time-redundant recomputation.

\item We begin our work by optimizing an FFT baseline without fault tolerance. Through a series of optimizations on tiling, thread-level workload balance, twiddling factor computation, and global memory access pattern, TurboFFT offers performance competitive to the state-of-the-art closed-source library, cuFFT. TurboFFT is available at an anonymous link.\footnote{https://github.com/ics2024artifact/ics2024artifact.git} 

% \item We develop TurboFFT, a lightweight yet high-performance FFT implementation. A state-of-the-art FFT implementation is the foundation for understanding the landscape of performance bottlenecks for a variety of input shapes, data types, and hardware specs, and developing our fault tolerance scheme efficiently. TurboFFT is capable of generating FFT kernels for NVIDIA GPUs, with a performance variation \textcolor{blue}{ranging from a 5\% decrease to a 5\% increase} compared to the closed-source state-of-the-art library, cuFFT. The related source codes of this paper are open-sourced at the anonymous link \footnote{https://github.com/ics2024artifact/ics2024artifact.git}.

\item We explore the fault tolerance FFT designs at the threadblock, and thread levels. The combination presents a low overhead even under hundreds of errors injected per minute. 

\item A template-based code generation strategy is developed to reduce development costs. The template-based can generate FFT kernels with or without fault tolerance for a wide range of input sizes and data types.

\item  Experimental results of single precision and double precision on an NVIDIA A100 server GPU and a Tesla Turing T4 GPU show that TurboFFT offers a competitive or superior performance compared to the state-of-the-art closed-source library cuFFT. TurboFFT only incurs a minimum overhead (7\% to 15\% on average) compared to cuFFT, even under hundreds of error injections per minute for both single and double precision. TurboFFT offers a 23\% improvement compared to existing fault tolerance FFT schemes.

\end{itemize}

\section{Background and Related Works}
\label{sec:background}

 % We present an analysis of two software strategies for tolerating runtime computing errors: duplication-based methods and algorithm-based approaches. In addition, we survey prior fault-tolerant FFT research that has employed the technique of kernel fusion, aiming to amplify the efficiency of applications, particularly those that utilize fast Fourier transform.
  FFT accelerates discrete Fourier transform (DFT) computation through factorization. The forward DFT maps a complex sequence $\mathbf{x}=(x_0, x_1,\cdots,x_{N-1})$ to $\mathbf{y}=(y_0,\cdots, y_{N-1})$, where 
$y_j = \sum_{n = 0}^{N-1} x_n e^{-2\pi ijn/N}$. The inverse discrete Fourier transform is defined as $x_j = \frac{1}{N}\sum_{n = 0}^{N-1} y_n e^{2\pi ijn/N}$. DFT can be treated as a matrix-vector multiplication (GEMV) between the DFT matrix $W=( \omega_N^{jk})_{j,k=0,\cdots,N-1}, \omega_N = e^{-2\pi i/N}$ and the input sequence $\mathbf{x}$. Given the expensive O($N^2$) operations of GEMV, FFT factorizes the DFT matrix into a sparse factors product and achieves a complexity of O($N\log N$) \cite{van1992computational}. In the following, we first present the fault model. Next, we describe previous works for FFT with or without fault tolerance.
 
\subsection{Fault Model}
% We focus on FFT for complex sequences using Stockhanm's FFT \cite{swarztrauber1984fft}.

% \subsection{Fault Model}
TurboFFT focuses on the detection and correction of errors at computing units that can affect the results of the final output. We assume memory errors are protected by ECC \cite{bird2017neutron} and the reliability issue of communication is protected by FT-MPI \cite{fagg2000ft}. To address the compute errors at the run-time,  we adopt a fault-tolerant scheme under a single-event upset (SEU) assumption \cite{binder1975satellite,petersen2013single,binder1975satellite}, i.e., there is only one soft error in each error detection and correction period. The SEU assumption is validated by the low occurrence rate of multiple soft errors caused by short fault detection intervals, widely used in many works \cite{reis2005swift, zhai2021ft, ding2011matrix, wu2014ft}. More specifically, a feasible fault tolerance scheme is capable of identifying injected faults with a high degree of reliability and a negligible false alarm rate in the following case \cite{turmon2000software}. Firstly, 2000 random test sequences are generated, and faults are injected in half of these runs (1000 of 2000) by first choosing an element to affect, and then flipping exactly one bit of its 32-bit representation for float-precision and 64-bit representation for double-precision. Secondly, a checksum test with threshold $\delta$ is used to attempt to identify the affected computations. Finally, an error correction scheme is applied.

\subsection{Previous Work for FFT without Fault Tolerance}

High-performance FFT on GPU is a crucial linear algebra routine for scientific computing, playing a significant role across various applications \cite{thompson2022lammps,kim2018qmcpack,habib2016hacc,stockman1999ngst, murphy1998ngst}. The popular FFT libraries, such as cuFFT 
 \cite{cufft}, rocFFT \cite{rocfft}, VkFFT \cite{tolmachev2023vkfft}, FFTX \cite{franchetti2018fftx}, and heFFTe \cite{ayala2020heffte} are designed explicitly for high-performance. cuFFT is one of the state-of-the-art closed-source libraries particularly designed for NVIDIA GPU \cite{cufft}. VkFFT, initially designed for Vulkan, supports CUDA/HIP/OpenCL as well \cite{tolmachev2023vkfft}. Motivated by exascale computing, FFTX adopts a code generation strategy to generate backend FFT kernels while heFFTe focuses on communication and leverages single GPU FFT kernels from cuFFT and rocFFT \cite{franchetti2018fftx,ayala2020heffte}. 
 
 The above FFT libraries are harnessed on top of a high-performance facility, fault tolerance should be orthogonal to the performance they provide. Our development of an FFT baseline from scratch is driven by the essential need to understand FFT performance bottlenecks, crucial for seamlessly integrating fault-tolerance operations. To keep aligned with the performance provided in those libraries, TurboFFT develops an FFT baseline competitive with one of the state-of-the-art library cuFFT. Although developed in CUDA, TurboFFT can be generalized to GPUs from other vendors as it is all programmed using high-level CUDA C++ and does not feature any arch-specific primitives such as PTX. We presume that vendor-provided programming language translation tools can port our library together with its parallel designs and optimization strategies to Intel DPC++/AMD HIP/OpenCL-based software.

\subsection{Previous Work for FFT with Fault Tolerance}
Existing fault-tolerant FFT attempts \cite{liang2017correcting,fu2009fault,pilla2014software,may1979alpha, baumann2002soft,geist2016supercomputing} focus on the ABFT schemes proposed in \cite{jou1988fault,wang1994algorithm}. We discuss the ABFT schemes and describe two implementations.
% Pilla \cite{pilla2014software} presents an offline implementation and Xin \cite{liang2017correcting} integrated ABFT into the FFT kernel, thereby reducing the overhead of ABFT and enabling it to correct errors before the computation finishes.

% Xin \cite{liang2017correcting} minimizes the ABFT overhead on CPU by fusing fault-tolerance instructions into FFTW \cite{liang2017correcting}. For GPU side, Pilla \cite{pilla2014software} first applies the ABFT scheme while. % In these methods, the FFT result is verified by an equation $(e^TW)\mathbf{x}=e^T(W\mathbf{x})$ and corrected with recompute.

\textbf{ABFT scheme.} Existing ABFT schemes detect errors by utilizing a perspective of GEMV in DFT, namely $(\mathbf{e}^TW)\mathbf{x}= \mathbf{e}^T(W\mathbf{x})$, as shown in Eqn. (\ref{eqn:ABFT}) and (\ref{eqn:C}).

\begin{equation}
    W\xrightarrow[]{encode}W^c:=\left[\begin{array}{c}W \\ \mathbf{e}^T W\end{array}\right],
    \label{eqn:ABFT}
\end{equation}

\begin{equation}
\mathbf{y}^c = W^c\cdot \mathbf{x} = \begin{bmatrix}W\mathbf{x} \\ \mathbf{e}^T W\mathbf{x} \end{bmatrix} = \begin{bmatrix}\mathbf{y} \\ \mathbf{e}^T\mathbf{y}
 \end{bmatrix},
\label{eqn:C}
\end{equation}
where $\mathbf{x}$ is the input signal, $X$ is the output signal, $W$ is the DFT matrix, and $\mathbf{e}$ is the encoding vector. The correctness is verified by comparing $(\mathbf{e}^TW)\mathbf{x}$ and $\mathbf{e}^T\mathbf{X}$. If $|(\mathbf{e}^TW)\mathbf{x} - \mathbf{e}^T\mathbf{X}|/|(\mathbf{e}^TW)\mathbf{x}|$ exceeds an error threshold $\delta$, it indicates that an error occurred during the computation. The selection of $\mathbf{e}$ is widely discussed because the 1's vector misses the opposite error $x+\epsilon$ and $x-\epsilon$ by addition. Jou \cite{jou1988fault} suggests $\mathbf{e}_{\text{Jou}} = (1/\omega_N^{-0},\cdots,1/\omega_N^{-(N-1)})$ while requiring a variant input $\mathbf{x}' = (2x_0+x_1,\cdots,2x_{N-1}+x_0)$, leading to additional computation overhead. Next, Wang \cite{wang1994algorithm} proposes $\mathbf{e}_{\text{Wang}} =  (\omega_3^0,\cdots,\omega_3^{N-1})$ allowing $\mathbf{x}$ to be unchanged. The computation of $\mathbf{e}^TW$ is not trivial and introduces addition computation or memory overhead.

Due to the error propagation, existing fault tolerance schemes not only necessitate a checksum computation for each signal and a time-redundant recompute under error correction. In contrast,  TurboFFT employs a column vector $\mathbf{e}_c$ to linearly combine a batch of signals, as shown in Eqn. (\ref{eqn:two_side_X}) and (\ref{eqn:two_side_Y}). The composite signal can recover the corrupted signal given the SEU assumption and the linearity of FFT, enabling batched detection and delayed correction without recomputation. 

% The ABFT in FFT is achieved by encoding discrete Fourier transform matrix $W$ into checksum forms $W^c$.  The encoded DFT matrix and the input signal $x$ are then multiplied together to produce a complex vector $X^f \in \mathbb{C}^{N+1}$ that contains both the correct result $X$ and checksum information $X^c \in \mathbb{C}$: 
% \begin{equation}
% B\xrightarrow[]{encode}B^r:=\begin{bmatrix}B & Be\end{bmatrix},
% \label{eqn:B_encode}
% \end{equation}

\textbf{Offline FT-FFT.} Pilla \cite{pilla2014software} adopts the encoding vector $\mathbf{e}_{\text{Jou}}$ and presents a software implementation on GPU. Although employing a parallel programming model, this work detects and
corrects errors until one FFT computation finishes, introducing high computation overhead. Besides, the implementation only applies to thread-level FFT due to the input remapping required by $\mathbf{e}_{\text{Jou}}$, neglecting the end-to-end FFT performance and higher-level hierarchy elements such as warp and threadblock. This work also lacks the integration of ABFT into a state-of-the-art GPU FFT library. For example, the variant input and the complicated encoding pattern of $\mathbf{e}_{\text{Jou}}$ will introduce high computation overhead to an FFT baseline, worsened by the expensive function calls of $\sin,\cos$ to obtain encoding vector. As opposed to that, TurboFFT not only implements an efficient FFT baseline competitive to cuFFT, but also presents a series of architecture-aware designs at threadblock, and thread level, maintaining an end-to-end overhead of 10\%. 
\begin{figure}[t!]
    \centering
    \includegraphics[width=\linewidth]{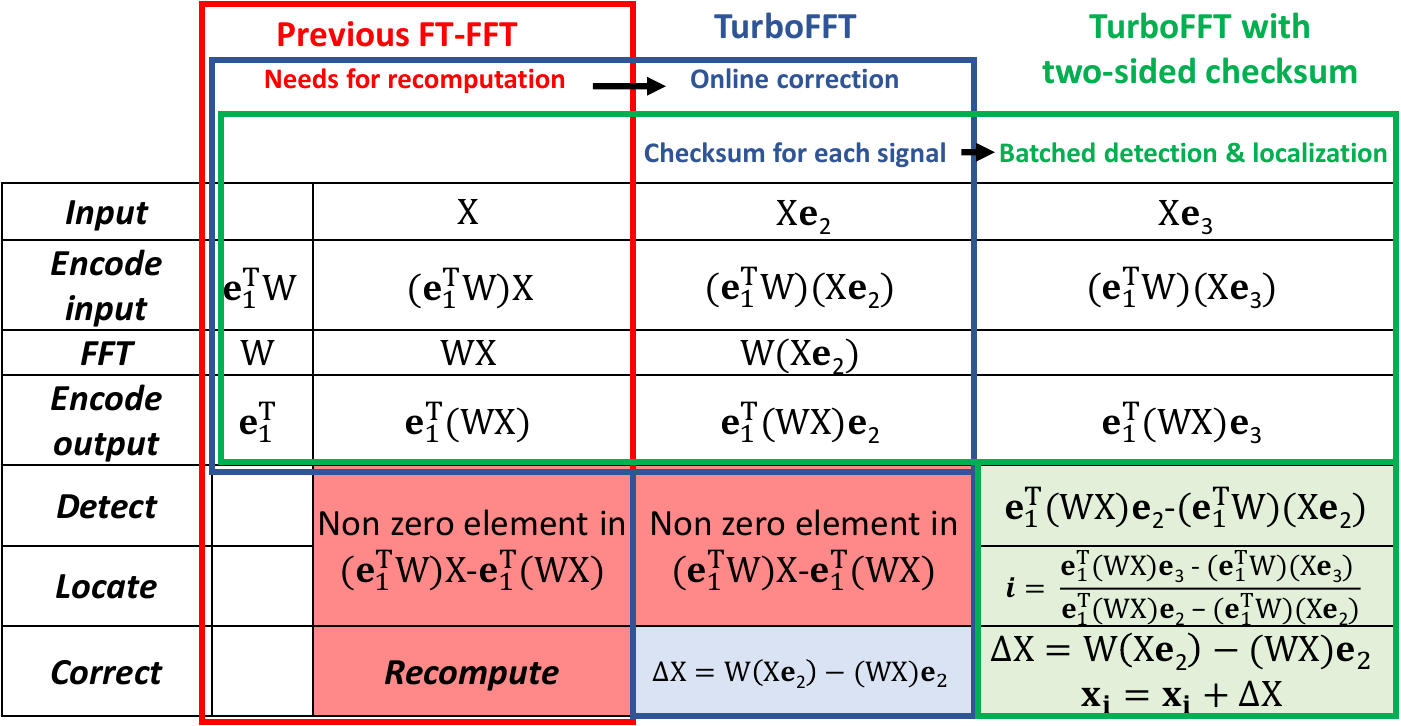}
    \caption{Motivation: Red area incurs high overhead. Blue region enables online correction. Green region enables batched detection.}
    \label{tab:illustrtation}
    \vspace{-1mm}
\end{figure}

\textbf{Online FT-FFT with kernel fusion.} Xin's FT-FFT \cite{liang2017correcting} presents a kernel fusion strategy for the CPU to hide the memory instruction overhead by integrating the ABFT operations into FFTW \cite{frigo1998fftw} kernels. Different from the offline FT-FFT, Xin adopts the encoding vector  $\mathbf{e}_{\text{Wang}}$ to simplify the encoding computation. Although Xin's FT-FFT achieves a $20\%$ overhead on CPUs, it not only requires a checksum computation for each signal per thread but also necessitates the computation of $\mathbf{e}^TW$. In Xin's implementation, $\mathbf{e}^TW$ is precomputed and loaded, allowing threads to fetch it from there as needed. However, GPU FFT is bouned by global memory transactions. This makes the ABFT operations of reading data from global memory extremely expensive, resulting in nearly $35\%$ overhead. To alleviate global memory transactions, our batched-detection approach effectively reduces global memory overhead in inverse proportion to the batch size per threadblock, while keeping the end-to-end overhead to around $10\%$.

 \begin{figure*}[ht]
     \centering
     \includegraphics[width=\linewidth]{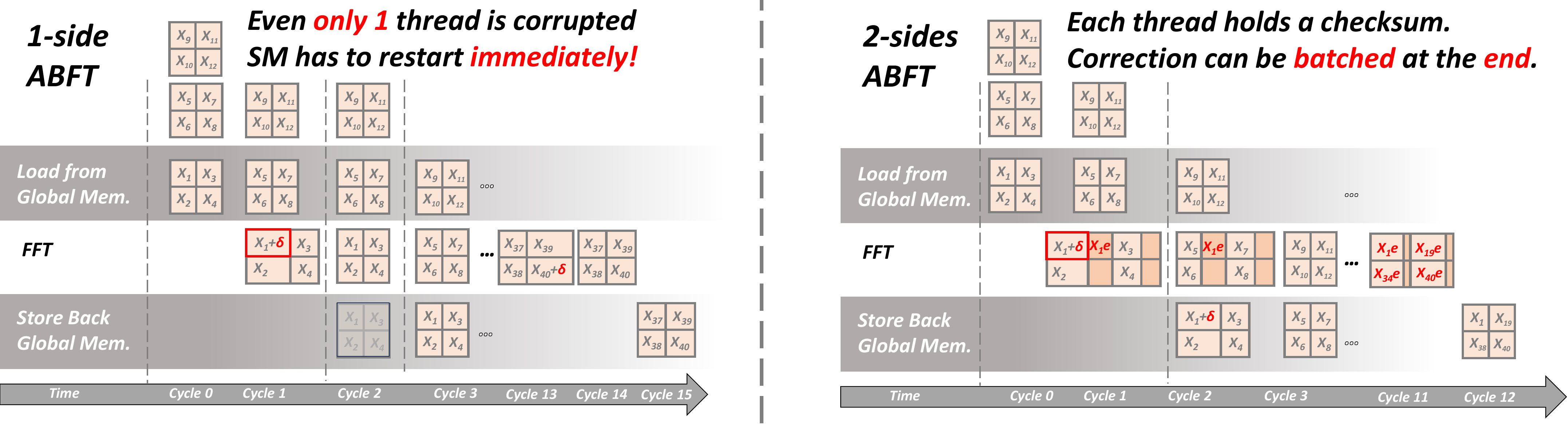}
     \caption{What is the benefit of two-sided ABFT? Delayed Batched Correction }
     \label{fig:why-two-sided-abft}
     \vspace{-3mm}
 \end{figure*}
\section{Two-sided Checksum}
Figure \ref{tab:illustrtation} details our motivation and algorithm designs. Previous FT-FFT, either offline or online, requires \textit{per-signal checksum}, namely applying the encoding vector to each input. Other than that, previous work states that a time-redundant recomputation is necessary. Motivated by the error propagation issue, we find the single error in FFT, although it corrupts hundreds of elements, the errors in the output are proportional to the initial error, as shown in the blue region in Figure \ref{tab:illustrtation}. To avoid frequent checksum encoding, we employ another encoding vector $\mathbf{e}_3=(1,2,\cdots,N)$ to \textit{linear combine} a batch of inputs within each thread and encode those thread-local variables with $\mathbf{e}_1$ at a fixed interval. The corrupted input can be located using a quotient shown in the green region of Figure \ref{tab:illustrtation}.

% \begin{table}[h]
% \centering
% \begin{tabular}{|p{0.11\linewidth}|p{0.25\linewidth}|p{0.06\linewidth}|p{0.30\linewidth}|p{0.11\linewidth}|}
% \hline
% \multirow{2}{*}{ABFT} & \multicolumn{2}{c|}{Error Detection} &\multicolumn{2}{c|}{Error Correction} \\ 
% \cline{2-5}
% & Computation & \#Ops & Computation & \#Ops  \\
% \hline
% \textcolor{blue}{1-sided} &  $(e^TW)X$,$e^T(WX)$ & $8n$ & $WX$ & $5n\log_2n$\\ 
% \hline
% 2-sided &  $(e^TW)X, W(Xe)$ $e^T(WX), (WX)e$ & $16n$ & $\Delta$$= W(Xe) - (WX)e$ $X\leftarrow X+\Delta$& $4n + 5n\log_2n$\\ 
% \hline
% \end{tabular}
% \caption{One-side ABFT vs. Two-sided ABFT}
% \label{table:alg}
% \vspace{-1mm}
% \end{table}

% We present our fault model before moving on to the fault tolerance design. This work focuses on the detection of errors in the computing units that can corrupt the results of the output signals. We assume the memory is protected by the ECC code. To robustly tackle these computational inconsistencies at run-time, we have devised a fault-tolerant scheme that operates under the assumption of a single-event upset (SEU). This assumption constrains the model to consider only one soft error in each error detection and correction period. The SEU assumption is not merely a theoretical convenience; it is a well-established and widely employed fault model in numerous research works \cite{reis2005swift, zhai2021ft, ding2011matrix, wu2014ft}. Its validity is corroborated by the empirically observed low occurrence rate of multiple soft errors, a phenomenon attributable to the brief intervals used in fault detection.

\subsection{Error Dectection}

 \begin{equation}
    X\xrightarrow[]{two-sided}X^c:=\begin{bmatrix}X &  X\mathbf{e}\end{bmatrix},
    \label{eqn:two_side_X}
\end{equation}
\begin{equation}
Y^c = W^c\cdot X^c = \begin{bmatrix}WX & WX\mathbf{e} \\ \mathbf{e}^T W\mathbf{x} & \mathbf{e}^TWX\mathbf{e} \end{bmatrix} = \begin{bmatrix}Y & Y\mathbf{e} \\ \mathbf{e}^TY & \mathbf{e}^TY\mathbf{e}
 \end{bmatrix},
\label{eqn:two_side_Y}
\end{equation}
There are two checksums being computed in the two-sided ABFT, namely the left-side $eWX$ and the right-side $WXe$. The checksum along the individual signal, namely one column in $X$, is used to detect the location of the error. Once a mismatch between the checksum before and after one FFT process, it is treated as an error. If an error is detected, then the divergence of the checksum along different signals between before and after the FFT is added back to the detected signal to perform the correction. According to Figure \ref{tab:illustrtation}, both two-sided ABFT and one-sided ABFT compute the checksum on the left side, i.e., eWX. In one-sided ABFT, this checksum is used to detect whether an error has occurred, and if an error is detected, the process reverts to a previously saved state for recalculation. In two-sided ABFT, while using this left-side checksum, the sum is also taken along the row direction of the current data X. If we perform an FFT operation on this checksum vector and subtract it from the checksum of the output result, we can then obtain the correction value for the entire column of erroneous data. Next, we just need to add this correction value back to correct the error.

 \subsection{Delayed Batched Correction}
The biggest difference between one-sided ABFT and two-sided ABFT is whether there is an immediate need for recalculation. From Fig. \ref{tab:illustrtation}, we understand that two-sided ABFT requires the calculation of an additional set of checksums, and correction also involves performing an FFT operation on the checksum vector. So, compared to one-sided ABFT, where does the advantage of two-sided ABFT lie? This work points to the advantage, namely delayed batched correction.

Under the assumptions of ABFT, a single checksum can correct one error. Therefore, for two-sided ABFT, it is only necessary to note the position of the error, i, and then continue processing the data for position i+1, until the operation ends or a new error occurs. In those cases, we need to correct the contents of the checksum at the erroneous position $i$. As a result, there is an opportunity for batch-processing operations among different threads, which enhances parallelism. Moreover, since there is no need to stop and execute immediately, the running pipeline of the program is not affected, thereby avoiding stalls.
 Figure \ref{fig:why-two-sided-abft} illustrates the difference between one-sided ABFT and two-sided ABFT. Any fault tolerance mechanism is essentially a tradeoff between resources. Compared to one-sided ABFT, two-sided ABFT actually uses additional computation to reduce memory overhead. If one-sided ABFT chooses not to recalculate immediately, it will need to reload data from the storage device during the next computation. In contrast, two-sided ABFT \textit{compresses} the data already read into local registers in the form of checksums. Therefore, before a new error occurs, the error information can be decompressed at any time by the checksum and used to correct previous errors. 

Fig. \ref{tab:illustrtation} demonstrates the capability of batch delay correction. The 2$\times$2 grid in the figure represents a warp. When thread 1 in a warp encounters an error, the entire warp can continue to operate normally until the program terminates or a new error occurs. In contrast, one-sided ABFT requires immediate re-execution, otherwise it would need to reload data from memory.

 Figure \ref{tab:illustrtation} details the computation process for two-sided ABFT. When an error occurs, a thread can note that an error has occurred at this point. Error correction operations are only carried out when the next error occurs or when the program terminates. Placing the right-side checksum at the end of the loop is to prevent a potential second error from contaminating the checksum that has already recorded one error.
% \begin{algorithm}
% \caption{Two-Sided ABFT FFT Algorithm}
% \begin{algorithmic}[1]
% \Procedure{Two-Sided-ABFT-FFT}{}
% \State Get initial radix $k$ and corresponding $m = \frac{N}{k}$
% \State Calculate left checksum vector $c_m = e_m W_m$
% \State Set the error flag $ErrFlag = \text{false}$
%     \State Set the error location $j = \text{-1}$
% \For{$i \text{ from } 0 \text{ to } k-1$}
%         \State Calculate the $i$-th FFT: $X'_i = W_m x_i$
%         \If {$(\lVert e_m X'_i - c_m x_i \rVert > \eta_1)$ or $i == k-1$}
%         \If {$ErrFlag$}
%         \State Correction $X'_j -= W_m r_m - p_m$
%         \EndIf
%         \State Set error flag $Errflag=True$
%         \State Record error location $j = i$
%         \EndIf
%         \State Calculate right checksum vector $p_m += X'_ie_m$
%          \State Calculate right checksum vector $r_m +=  x_i e_m$
% \EndFor
% \EndProcedure
% \end{algorithmic}
% \end{algorithm}

\section{Design and Optimizations}
\label{sec:design_and_optimizations}
A high-performance fault-tolerant FFT implementation necessitates fusing the memory costs of fault detection with the original FFT operations. Hence, an efficient FFT framework serves as the foundation for memory operations of fault detection. As cuFFT is closed-source, implementing a high-performance FFT from scratch becomes inevitable. In this section, we present the step-wise optimizations of FFT. The optimizations include avoiding bank conflict, maximizing the L1 cache hit rate, reducing triangular operations, and a template-based code generation for parameter selection.
\subsection{TurboFFT Stepwise Optimization}
% \subsubsection{Basic implementation}
% As the basic version, we implement TurboFFT only considering application level, as shown in Figure 4. To simplify the development, we load data into registers through shared memory if applicable. Otherwise, the input signal is loaded into registers from globally directly. The memory traffic is directly through global memory to registers. Each thread holds two complex elements. To better utilize the memory locality, Stockham version Cooley-Tukey FFT is used. In Stockham algorithm, consecutive threads $i, i+1$ perform two adjacent radix-2 FFTs, namely FFT($x[i], x[i+N/2]$) and FFT($x[i+1], x[i+1+N/2]$), where $N$ is the raw FFT size. After the radix-2 FFT is finished by threads, a twiddling stage with global index is performed. After that, the result will be stored in global memory. The global memory access still obeys thread-level locality so that the store operations are coalesced between consecutive threads. Because of the lack of synchronization techniques between different threadblocks, multiple kernel launches are required if the application level FFT size cannot fit into the shared memory. Hence, the basic TurboFFT kernel is called $log_2N$ times and introduces a high overhead for an FFT size larger than $8192$. The performance ratio between basic version TurboFFT and cuFFT is $4.5\%$.
\subsubsection{Tiled FFT}
Figure \ref{fig:turbo_fft_details} demonstrates the tiled FFT at the kernel level. When the FFT size cannot fit into a threadblock's shared memory, e.g. 64KB for T4, and 192 KB for A100, multiple kernel launches are inevitable. The original FFT signal with size $N$ is divided into multiple smaller tiles, $N_1\times N_2 \times N_3$, so that each segment can be independently processed within the shared memory of a threadblock. Through multiple kernel launches, these tiles are processed sequentially, and the results are combined when writing back. 
% Tiled FFT allows for the effective processing of FFTs larger than the size of the shared memory.
% As shown in Figure \ref{fig:turbo_fft_details}, the application FFT signals are rearranged into a 3D cube in kernel level without data exchange. Hence, the tiling operation introduces no additional overhead while only requiring a perspective transfer from 1D to 2D or 3D in data layouts. By applying the tiling transform, each kernel solves a batch of small-sized FFT ($N \leq 8192$) in one kernel launch, which greatly minimizes the memory cost in multiple launches. We assume the 3D cube obeys an index order of $(N_1, N_2, N_3)$, where the adjacent elements from the outer-most axis $X_1$ have a stride of $N_2\cdot N_3$, the adjacent elements from axis $X_2$ have a stride of $N_3$, and adjacent elements from axis $X_3$ has a stride of one. For the kernel launches along axis $N_1$ and $N_2$, no data transpose is introduced while the last kernel launch along axis $N_3$ transposes the element $(x_1, x_2, x_3)$ into $(x_3, x_2, x_1)$. In other words, the first two launches store the results to exactly the same indexes at outputs as inputs. Exploiting the data tiling, the performance boosts to a performance ratio of $55.5\%$ compared to cuFFT.

\subsubsection{Thread-level workload assignment}
The single batched radix-2 FFT only requires two complex number additions resulting in a highly underutilized computation capability, especially for the application-level FFT sizes smaller than $8192$  To increase the workloads of each thread, we allocate $8$, $16$, or $32$ elements for each thread. We integrate the thread-level FFT into a macro kernel, which takes thread-level FFT signals with input. FFT size less than $2^{13}$ benefits most from the increment of elements per thread, which obtains a performance ratio of $70.9\%$ when compared to cuFFT.

\subsubsection{Twiddling Factor Optimization}
Despite FFT's memory-bound nature, the Fast Fourier Transform (FFT) becomes compute-bound when the thread-level radix is $32$ or larger, especially for the twiddling factor computation. The twiddling factor computation incurs high overhead due to the high processor cycles used by $\sin(x)$ and $\cos(x)$. In our basic FFT kernels, there are 3 levels of twiddling stage, which are thread-level, warp-level, and threadblock-level. For thread-level, the twiddling factor occupies no more than 32 elements. Hence the twiddling factor can be encoded as constant into the thread-level macro FFT kernel. For warp-level, all $\sin(x)$ and $\cos(x)$ function calls can be replaced with complex number multiplications except for the first call. The thread-block level twiddling factors are obtained using two different ways for single precision and double precision. For single precision, the twiddling factors are obtained by function calls. For double precision, the overhead of $\sin(x)$ and $\cos(x)$ can occupy up to 20\% of the total execution time, hence the twiddling factors are prepared in global memory to be fetched. By reducing the number of trigonometric function calls, TurboFFT obtains a performance ratio of $80.9\%$ on average compared to cuFFT.

\begin{figure}[t]
    \centering
    \includegraphics[scale=0.14]{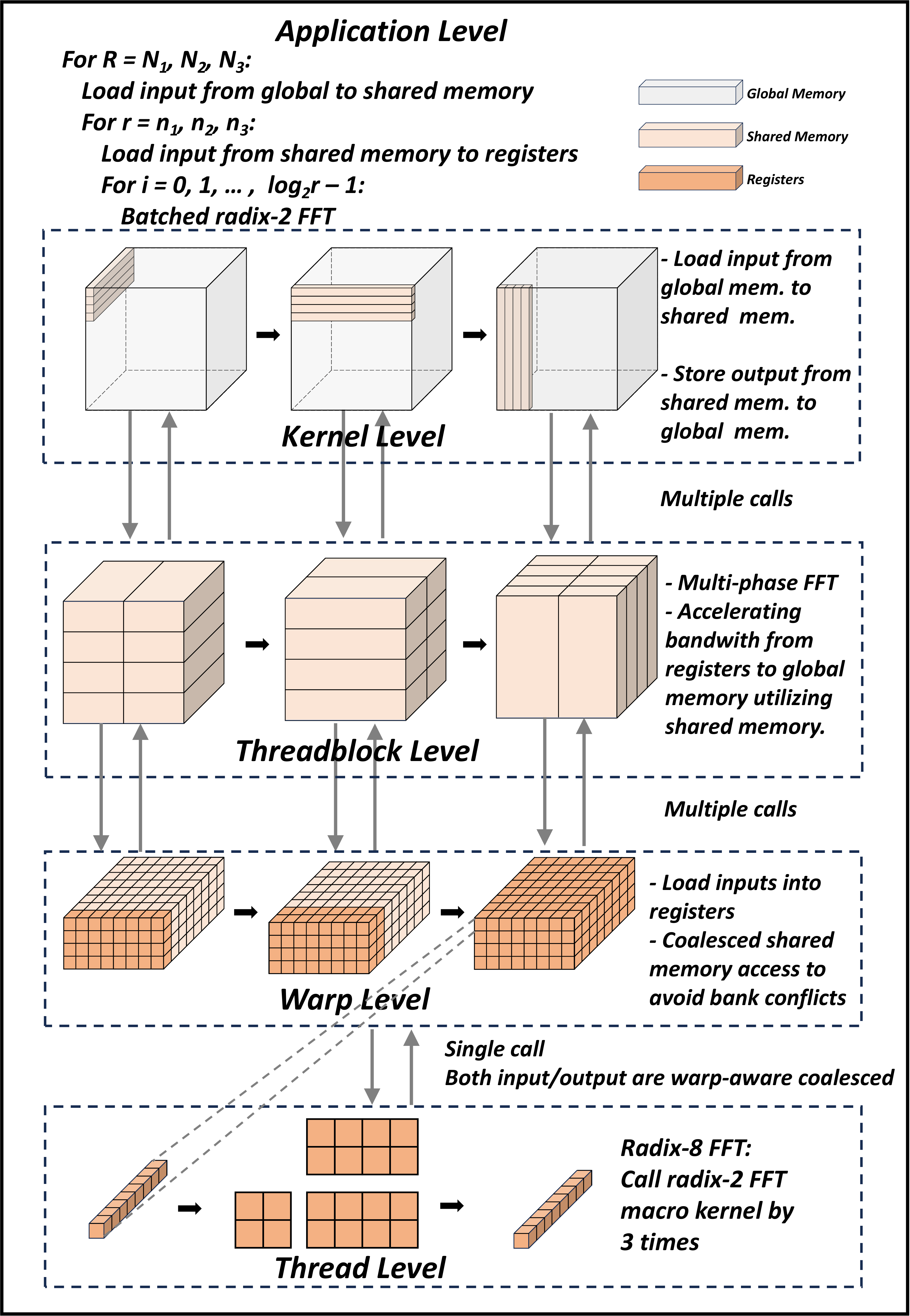}
    \caption{Overview of the optimized TurboFFT.}
    \vspace{-2mm}
    \label{fig:turbo_fft_details}
\end{figure}

\subsubsection{Global memory access pattern}
Although the efficiency has been greatly improved, our TurboFFT still suffers from a high L1-cache miss rate for tiled 3D FFT. Although the global memory access has greatly coalesced in warp-level with the Stockham algorithm, scattered global memory access is inevitable for the third FFT kernel launch in tiled 3D FFT. The scattered access will have a stride around $2^{20}$ due to the transpose operation, which incurs great inefficiency in the memory access pattern. To address the scattered global memory access pattern, TurboFFT assigns plane $N_1\times N_3$ to each threadblock at the last kernel launch, instead of plane $N_1 \times N_2$. By optimizing global memory access patterns, the performance ratio of TurboFFT, reaches a performance ratio of $98.5\%$ compared to cuFFT.

Figure \ref{fig:turbo_fft_details} demonstrates an overview of our optimized TurboFFT pipeline for FFT with large inputs, utilizing the step-wise optimizations mentioned above. Starting from the application level, the input signal is first tiled into a $N_1\times N_2\times N_3$ cube to fit into the shared memory maximum size of a threadblock. Next in the kernel level,  three stages of FFTs are performed step by step along each axis. For each stage, the FFT workload will be executed by threadblock, and each threadblock is assigned to a batch of FFT along the axis. Then at the threadblock level, the batched FFT loaded from global memory will first be arranged into a cube as the kernel level did. The cube size is equal to the thread-level radix. After that, warp-level FFT will load the input from shared memory into thread-level registers layer by layer. The shared memory access is coalesced to avoid bank conflicts and maximize the shared memory bandwidth. Once the thread-level inputs have been loaded into registers, the thread-level FFT macro kernel will perform the FFT computations. Within each level, the workflow goes from left to right. If a particular level's workload is completed, the tasks are then pop back to the higher level.

% \begin{table}[h]
% \centering
% \begin{tabular}{|p{0.2\linewidth}|p{0.25\linewidth}|p{0.15\linewidth}|p{0.25\linewidth}|}
% \hline
% \multirow{2}{*}{Method} & \multicolumn{2}{c|}{Error Detection} &\multirow{2}{*}{Error Correction} \\ 
% \cline{2-3}
% & Computation & Memory &  \\
% \hline
% Banerjee \cite{pilla2014software} &  One-side ABFT & ECC & Recomputation \\ 
% \hline
% Xin \cite{liang2017correcting} & One-side ABFT& Checksum & Recomputation \\ 
% \hline
% \textbf{Ours}& \multicolumn{2}{c|}{Two-sides ABFT}& On-the-Fly \\ \hline
% \end{tabular}
% \caption{Fault Tolerance Schemes with Architecture-Aware Optimizations}
% \label{table:FT_kernel_fusion}
% \vspace{-5mm}
% \end{table}

% \begin{figure}[t]
%     \centering
%     \includegraphics[width=1\linewidth]{figs/offline_overhead.pdf}
%     \caption{Offline ABFT vs. cuFFT: Offline ABFT has a significant overhead (up to 100\%) for small FFT sizes and batch sizes and an overhead of around 20\% for large FFT sizes and batch sizes. 2-sides ABFT holds a 10\%-20\% more overhead compared to 1-side ABFT. }
%     \label{fig:offline_overhead}
%     \vspace{-5mm}
% \end{figure}

\subsection{Optimizing TurboFFT With Fault Tolerance}
\label{sec:fault_tolerant}

The lightweight, high-performance TurboFFT provides us with an efficient framework to build lightweight fault-tolerant schemes from silent data corruption. This section introduces how we optimize the two-sided ABFT schemes in 4 steps. In summary, we first implemented an offline version based on cufft and cuBLAS. We then realized that SGEMV requires a single thread to traverse all batches or every element of a single signal. As shown in Figure \ref{tab:illustrtation}, offline ABFT, due to the need for checksum of all the data, actually doubles the memory transactions. Therefore, the overhead for both one-sided and two-sided ABFT is very close to 100\%. To avoid such a huge memory overhead, we decided to reduce the workload of each thread. This meant dividing the original signal into certain batch sizes, so that each thread's task shifted from computing the checksums of all batches to just a small part of them. Next, we connected our customized kernel with TurboFFT. However, the additional checksum still brought significant extra memory access. Ultimately, we found that we could fully fuse the ABFT checksum within a single thread, thereby reducing the additional memory operations to zero. Since the initial steps merely involve calling library functions and are quite straightforward, we proceed directly to the discussion of kernel fusion.

\begin{figure}[t]
    \centering
\includegraphics[width=0.9\linewidth]{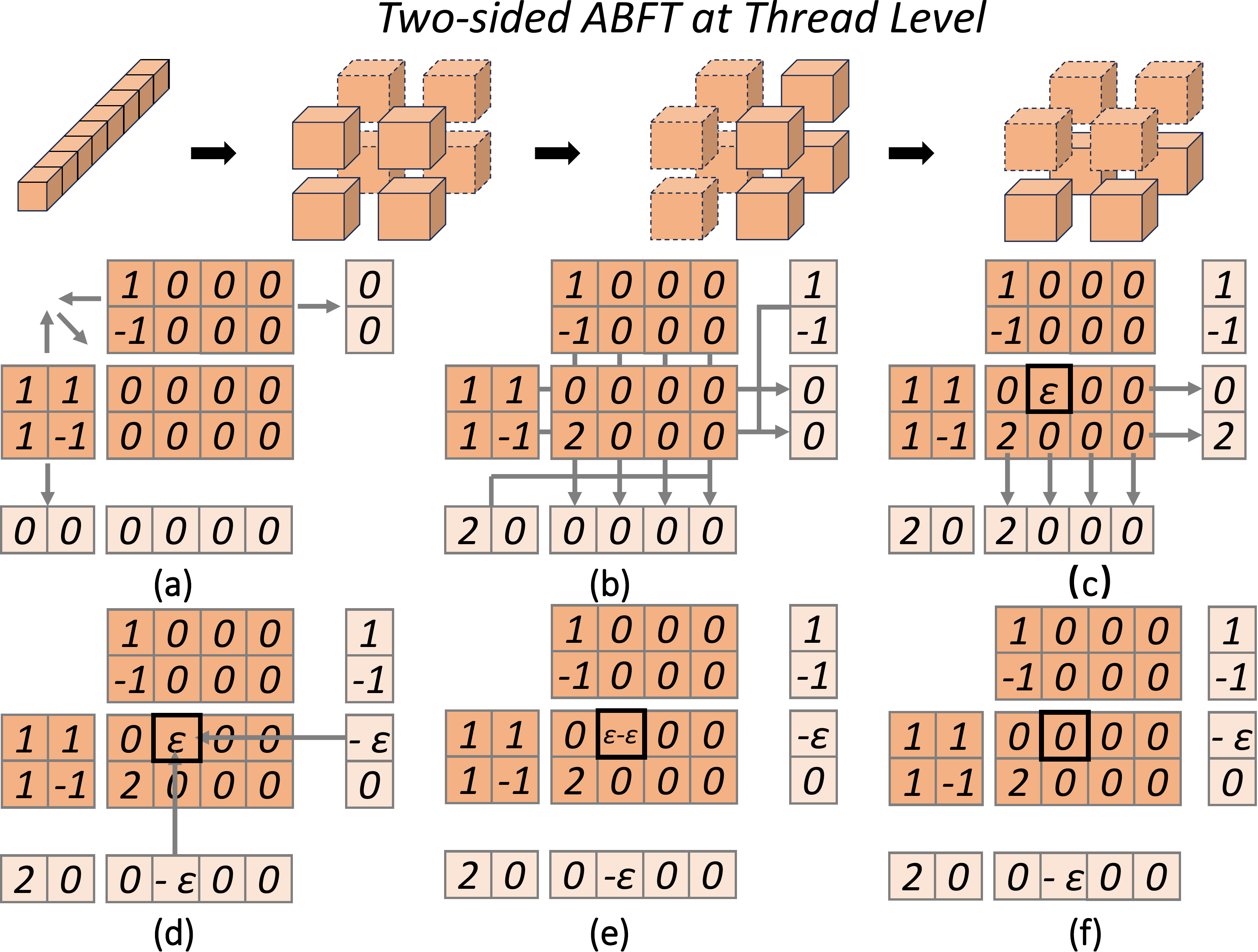}
    \caption{Thread-level Two-sided ABFT.}
    \label{fig:cft}
    \vspace{-4mm}
\end{figure}
\subsubsection{Thread-level Two-sided ABFT}

In Figure \ref{fig:cft}, we present how to use the two-sided ABFT to protect TurboFFT in thread level. As shown in Figure \ref{fig:cft} (a), right-side checksums of the radix-2 DFT matrix and input signals are encoded while the batched FFT is being computed. Next in Figure \ref{fig:cft} (b), the row and column checksums of output signals are computed through the matrix-vector multiplication between DFT matrix/input signals and their encoded checksums. If there is an error $\epsilon$ in Figure \ref{fig:cft}(c), then the row and column checksums of output signals from reduction will hold a disagreement of $-\epsilon$ with the previous checksums. After that, the error can be further located in Figure \ref{fig:cft}(d)  and corrected with the disagreement value in Figure \ref{fig:cft}(e). Utilizing the ABFT scheme, the in-register computation will be protected from silent error. Figure \ref{alg:codegen} demonstrates the pseudocode of the computation fault-tolerant scheme. Although it does not introduce additional memory overhead, there is excessive redundant computation among different threads, leading to significant overhead.

% \begin{figure}[ht]
%     \centering
%     \includegraphics[scale=0.22]{figs/c_pesudocode.pdf}
%       \caption{Hard-coded pseudocode: TurboFFT vs TurboFFT w/ FT. The ABFT-related operations are marked in red.}
%     \label{alg:c}
%     \vspace{-1mm}
% \end{figure}

\subsubsection{Threadblock-level Two-sided ABFT}
Compared to thread-level ABFT, threadblock-level ABFT can distribute the same checksum workload among different threads, thereby spreading out the extra checksum overhead. Although Liang et. al \cite{liang2017correcting} and Pila \cite{pilla2014software} both proposed kernel fusion strategies, neither is efficient enough. Their protection scheme requires multiple upload/download stages between memory and registers, exacerbating the extreme demand for memory bandwidth in GPU FFT workflows. Hence, we minimize the memory bandwidth occupancy of fault tolerance-related operations by utilizing warp shuffling, a synchronization-free instruction to exchange value between different threads in the same warp.  In the following, we detail the threadblock-level two-sided ABFT.

\begin{figure}[t]
    \centering
\includegraphics[scale=0.16]{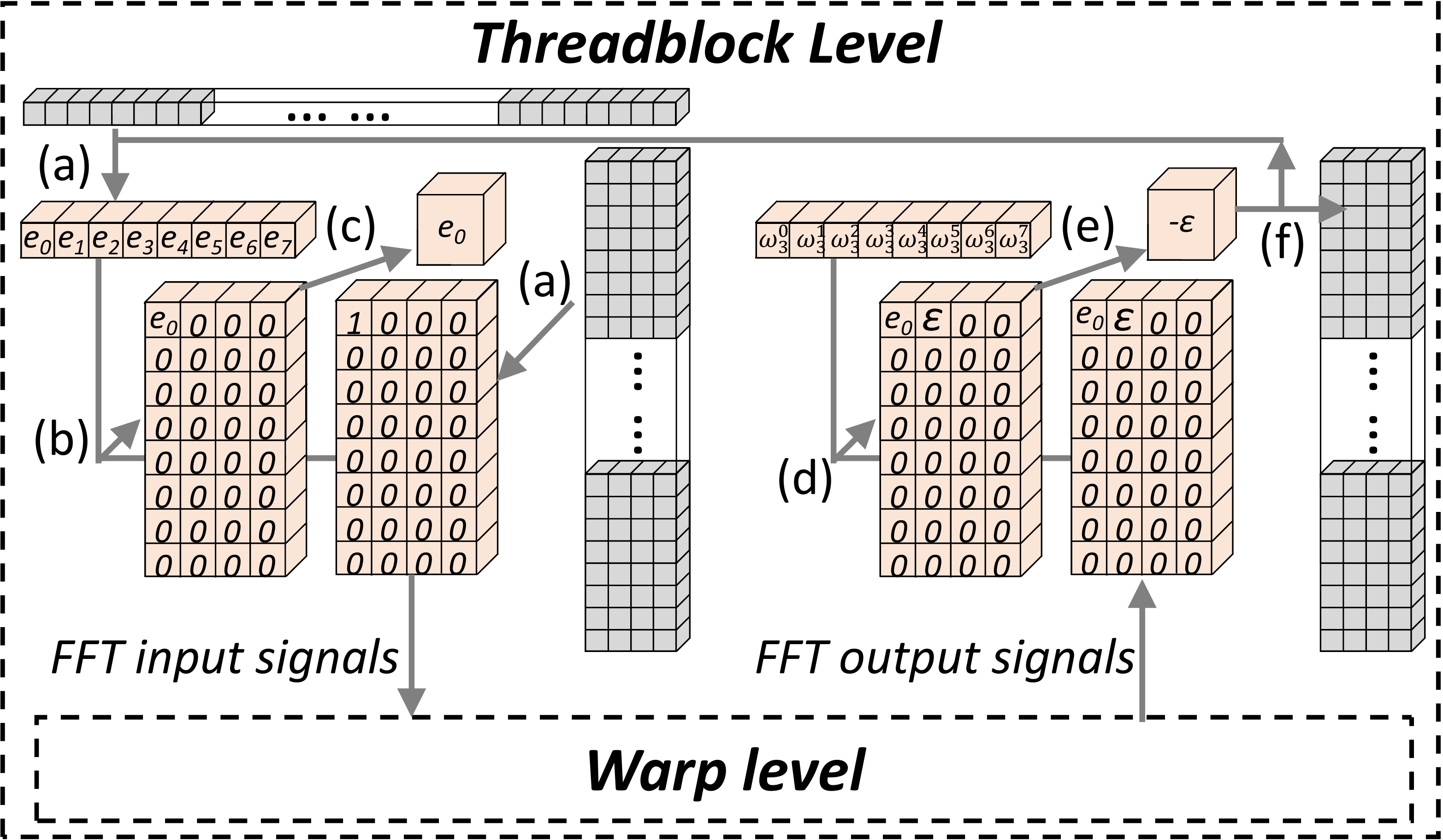}
    \caption{Checksum encoding at Warp Level.}
    \label{fig:s}
    \vspace{-3mm}
\end{figure}

We protect the FFT at the threadblock level and warp level, as shown in Figure \ref{fig:s}. In TurboFFT, the input of threadblock-level FFT is batched of signals, and then each thread first perform the right-side ABFT, namely a vector addition, where the vector is the input signals from the global memory. No additional memory transaction is required. The encoding vector of the left-side checksum $e^TW$ is precomputed outside the FFT application and loaded into the threadblock from global memory into the shared memory, as shown in \ref{fig:s} (a). The input signal encoding is performed through register reuse, which is in conjunction with loading input signals from global memory. After that, the DFT checksum is obtained using CUDA warp shuffle primitives. Each thread then keeps a checksum of the DFT result in Figure \ref{fig:s} (b), (c). Next, the register stores the original inputs into the shared memory to wait for the execution of warp-level FFT. When the output signals are returned from warp level, one ABFT encoding is performed again to verify the correctness of the output signals. The access to the encoding vector does not occupy additional memory bandwidth while only requiring a minimal overhead of warp shuffling, as shown in Figure \ref{fig:s} (d), (e). If an error is detected in Figure \ref{fig:s} (f), the corresponding thread will record the location (batch ID of the input signal). Then the output signals will be uploaded back to global memory for subsequent workloads.

\begin{figure}[thp]
    \centering
    \includegraphics[scale=0.22]{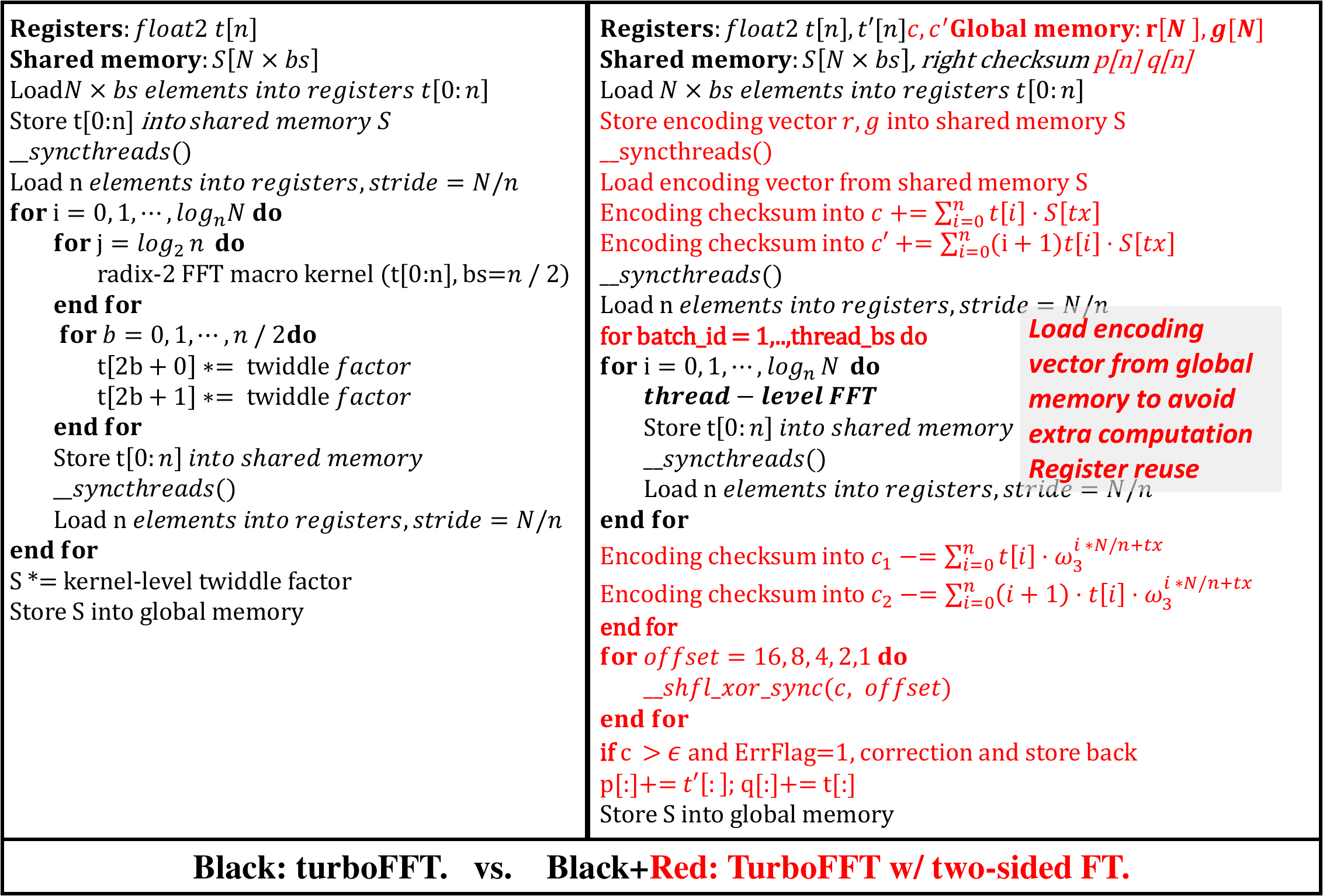}
      \caption{Codegen pseudocode: TurboFFT vs TurboFFT w/ FT.}
    \label{alg:codegen}
    \vspace{-2mm}
\end{figure}

\subsubsection{Automatic Code Generation}
A hard-coded FFT kernel can degrade performance when applied to different input sizes. However, rebuilding a FFT kernel with different parameter settings from scratch each time is not practical due to the high development cost. Generally, each FFT kernel in TurboFFT has around 2000 to 3000 lines of code (LOC) while a FFT with a large input size requires 3 different FFT kernels as mentioned above. To mitigate the increment of development cost, we propose a template-based code generation scheme to generate a series of input-size specific kernels. The code generation strategy we've developed involves utilizing semi-empirically chosen kernel parameters, tailored to input shapes, to generate highly efficient, parameterized kernels in real time. This dynamic code generation method enables us to extend the exceptional performance traditionally achieved by FFT on large sizes to different input and batch sizes, all the while maintaining minimal development costs.
% The static hard-coded parameter selection mechanism can degrade performance when applied to different input sizes. However, the development cost of a new parameter settings is extremely expensive. For instance, the TurboFFT kernel of a fixed parameter settings. However, the same parameters result in computing resources being underutilized because of the insufficient active threadblocks during runtime for smaller FFT input and batch sizes. To reduce the , we present a template-based code generation scheme. However, integrating a series of hard-coded kernels incurs significant development costs. As such, we utilize a templatized approach for our SGEMM kernels, which involves feeding semi-empirically selected kernel parameters according to the input shapes that generate high-performance parameterized kernels at runtime. This online code generation scheme allows us to generalize the superior performance of SGEMM on square matrices to irregularly shaped inputs while keeping development costs low.

\begin{table}[ht] \centering
%\vspace{-1mm}
\caption{TurboFFT kernel parameter setup on Tesla T4.}
\begin{tabular}{l
    S[table-format=3] % Align column by 3-digit int
    S[table-format=3] % Align column by 3-digit int
    S[table-format=3] % Align column by 4-digit int
    S[table-format=3] % Align column by 4-digit int
    S[table-format=3] % Align column by 4-digit int
    S[table-format=3] % Align column by 4-digit int
    S[table-format=3]
    }
\toprule
            {$N$}        & {$N_{1}$}       & {$N_{2}$}       & {$N_{3}$}  & {$n_1$} & {$n_2$} & {$n_3$} & {$bs$} \\ 
\midrule
$2^{10}$              & {$2^{10}$}          & {$$}              & {$$}    & {$8$} & {$$}    & {$$}   &  {$1$} \\ 
$2^{17}$              & {$2^8$}         & {$2^9$}             & {$$}      & {$16$} & {$16$}    & {$$} & {$8$}\\ 
$2^{23}$    & {$2^8$}         & {$2^7$}             & {$2^8$}      & {$16$}  & {$16$}   & {$16$} &{$16$} \\  
\bottomrule
\end{tabular}
\label{tab:kernel_size}
\vspace{-4mm}
\end{table}

\textit{Code Generation Strategy} Besides following the step-wise TurboFFT optimization, the code generation scheme takes 7 parameters as input and generates a corresponding high-performance FFT kernel. The kernel parameters are $N_1, N_2, N_3, n_1, n_2, n_3$ and $bs$. $N_1, N_2, N_3$ are the cube size of the kernel-level input signal. $n_1, n_2, n_3$ are the cube size of the threadblock-level input signal. $bs$ denotes the number of FFT signals a thread will take for one computation. In the code generation template, the memory operations are strategically designed to sidestep bank conflicts. Additional parameters, such as the data type for vectorized load/store operations, are determined directly by threadblock-level input sizes and batch size. Figure \ref{alg:codegen} shows the code generation template of FFT.

\textit{Kernel Parameters}
We generated FFT kernels with various parameters using the code generation template. Through empirical analysis, we identified a series of best kernels for input shapes from $2^3$ to $2^{29}$ and batch size from $1$ to $1024$, respectively. For input size in  $0\sim2^{13}$, $2^{14}\sim2^{22}$, and $2^{23}\sim2^{29}$, we adopt one, two, and three FFT kernel launches, respectively. Table \ref{tab:kernel_size} demonstrates the kernel parameter setting of 3 different kernels.

\section{Performance Evaluation}
\label{sec:results}

We evaluate TurboFFT on two NVIDIA GPUs, a Tesla Turing T4 and a 40GB A100-PCIE GPU. The Tesla T4 GPU is connected to a node with two 16-core Intel Xeon Silver 4216 CPUs, whose boost frequency is up to 3.2 GHz. The associated CPU main memory system has a capacity of 512 GB at 2400 MHz. The A100 GPU is connected to a node with one 64-core AMD EPYC 7763 CPU with a boost frequency of 3.5 GHz. We compile programs using CUDA $\mathtt{11.6}$ with the $\mathtt{-O3}$ optimization flag on the Tesla T4 machine, and using CUDA $\mathtt{12.0}$ on the A100 machine. A100 has a peak computational performance of $19.5$ TFLOPS for single precision and $9.7$ TFLOPS for double precision. The memory bandwidth is $1.55$ TB/s. T4 has a peak performance of $8.1$ TFLOPS for single precision and a peak performance of $0.253$ TFLOPS for double precision. The bandwidth of T4 is $320$ GB/s. We first demonstrate the benchmark result between TurboFFT and cuFFT for FP32, and FP64 FFT on A100. Next, we demonstrate the thread-level and threadblock-level design according to the proposed two-sided ABFT scheme. Finally, we present a fault coverage experiment and benchmark TurboFFT using two-sided ABFT under error injections with cuFFT, and the one-sided ABFT schemes presented in \cite{liang2017correcting,pilla2014software}. All experimental results are averaged over ten trials.

\subsection{Benchmarking TurboFFT without Fault Tolerance}

\begin{figure}[tp]
    \centering
    \includegraphics[width=\linewidth]{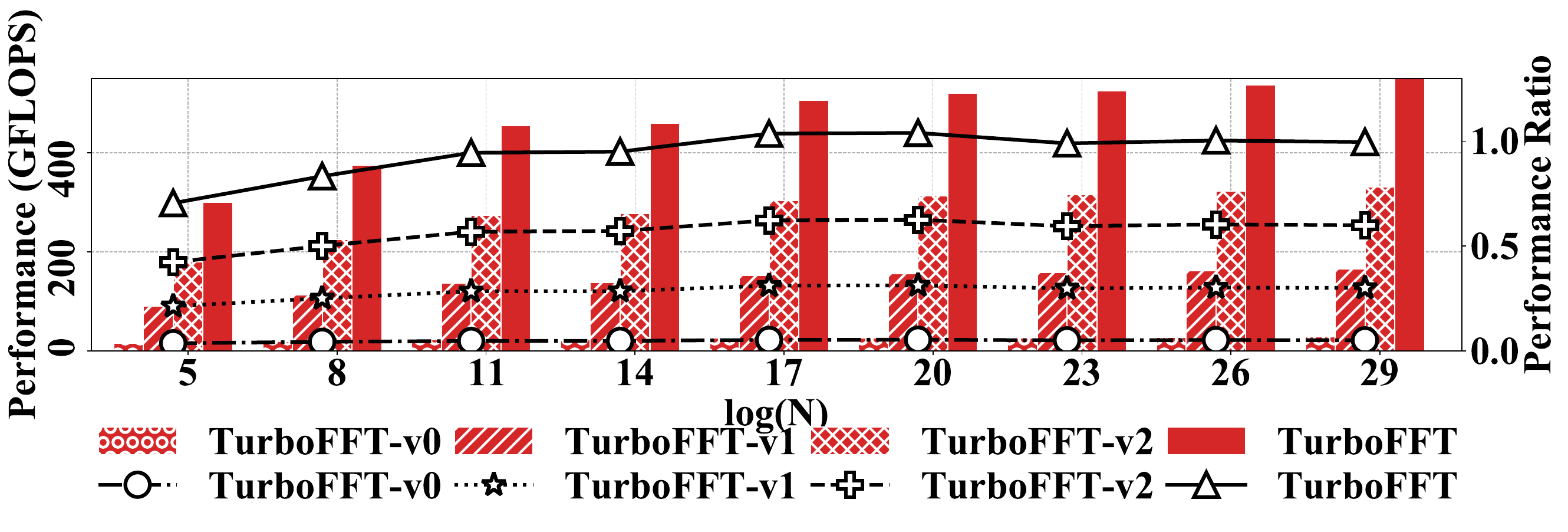}
    \caption{TurboFFT w/o FT stepwise optimizations on T4, FP32}
    \label{fig:stepwise_FP32_t4}
    \vspace{-1mm}
\end{figure}
\subsubsection{Step-wise optimizations for TurboFFT} Figure \ref{fig:stepwise_FP32_t4} presents the step-wise optimizations of TurboFFT without fault tolerance.  Figure \ref{fig:bench_FP32_a100} demonstrates how our TurboFFT kernel is optimized from 3\% to 99\% of cuFFT stepwise. The performance is measured with GFLOPS (bar plot, the left y-axis) and the performance ratio with respect to cuFFT (line chart, the right y-axis). For the most basic version, TurboFFT-v0, each thread handles a radix-2 FFT, and a $log_2(N)$ times of kernel launches are required. Without using any optimizations, the TurboFFT-v0 obtains a performance of 49 GFLOPS. Next TurboFFT-v1 employs the tiling strategy to split the original FFT into two or three subtasks. Each task is a batched FFT computation with a smaller FFT size. The number of kernel launches and subtasks are equal. The performance improved from 49 GFLOPS to 110 GFLOPS. Then, TurboFFT-v2 reconfigures the workload 
 assigned to each thread, e.g., increase the thread-level FFT size from $2$ to $16$, and optimize the twiddling factor computations. The performance is improved to 334 GFLOPS. After that, we optimize the memory access pattern, e.g. a threadblock uses a stride of $N_1\times N_2$ instead of $N_1$. The performance of TurboFFT achieves 565 GFLOPS. By employing the above strategies, we finally obtained an FFT baseline comparable to cuFFT. 

\begin{figure}[h]
    \vspace{-3mm}
    \centering
    \includegraphics[width=\linewidth]{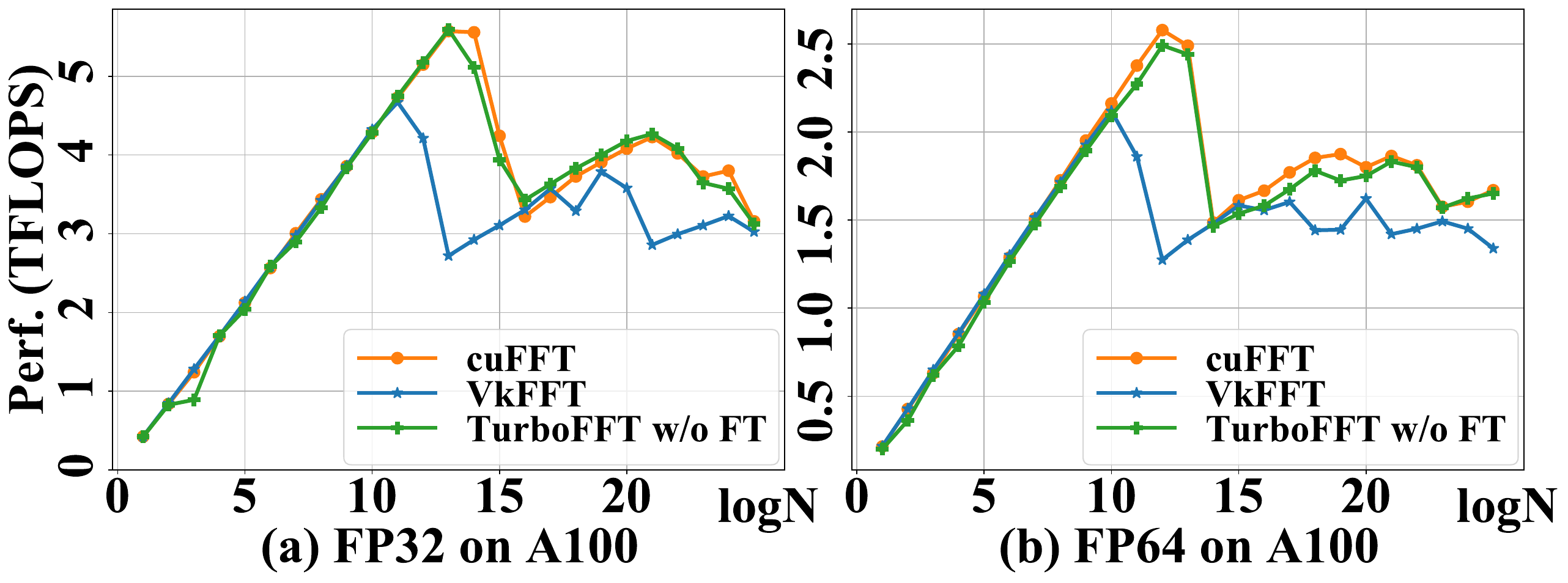}
    \caption{Comparison of batched FFT performance without fault tolerance with TurboFFT, cuFFT and VkFFT on an A100 GPU.}
    \label{fig:a100_bench}
    \vspace{-3mm}
\end{figure}

Figure \ref{fig:a100_bench} demonstrates a a performance evaluation between VkFFT and TurboFFT (without fault tolerance) relative to cuFFT. For FP32 precision, vkFFT and TurboFFT (without fault tolerance) show different levels of overhead compared to NVIDIA's cuFFT. VkFFT has an average overhead of 10\%, but can perform up to 3\% better or 51\% worse than cuFFT in certain scenarios. TurboFFT, on the other hand, presents a minimal average overhead of 2\%, with potential performance gains of up to 7\% in the best cases and a maximum overhead of 28\% in the worst cases. For double precision, VkFFT has an average overhead of 11\% compared to cuFFT, with performance varying between 3\% faster and 51\% slower. TurboFFT, on the other hand, has a lower average overhead of 4\%, with its performance ranging from 1\% faster to 15\% slower than cuFFT. TurboFFT provides more consistent performance closely aligned with cuFFT's. The performance of VkFFT drops at $\log N= 13,14$ due to an unbalanced workload for each thread. More specifically, VkFFT assigns a thread-level FFT with size 32, resulting in a computation bottleneck and waste of memory throughput.  
\begin{figure}[th]
\vspace{-2mm}
    \centering
    \includegraphics[width=1\linewidth]{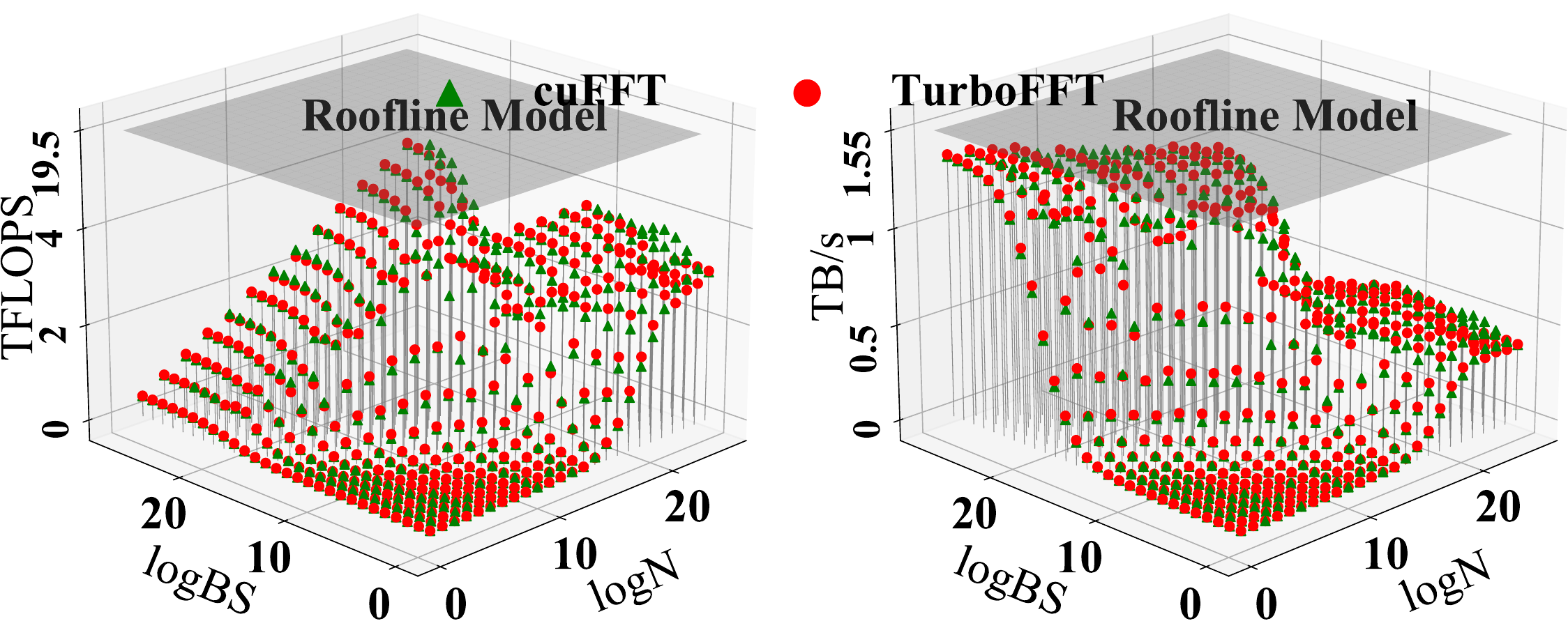}
    \caption{Performance of generated FP32 FFT kernels on A100}
    \label{fig:bench_FP32_a100}
    \vspace{-2mm}
\end{figure}

\subsubsection{Automatic code generation} 
\begin{figure}[hp]
\vspace{-3mm}
    \centering
    \includegraphics[width=1\linewidth]{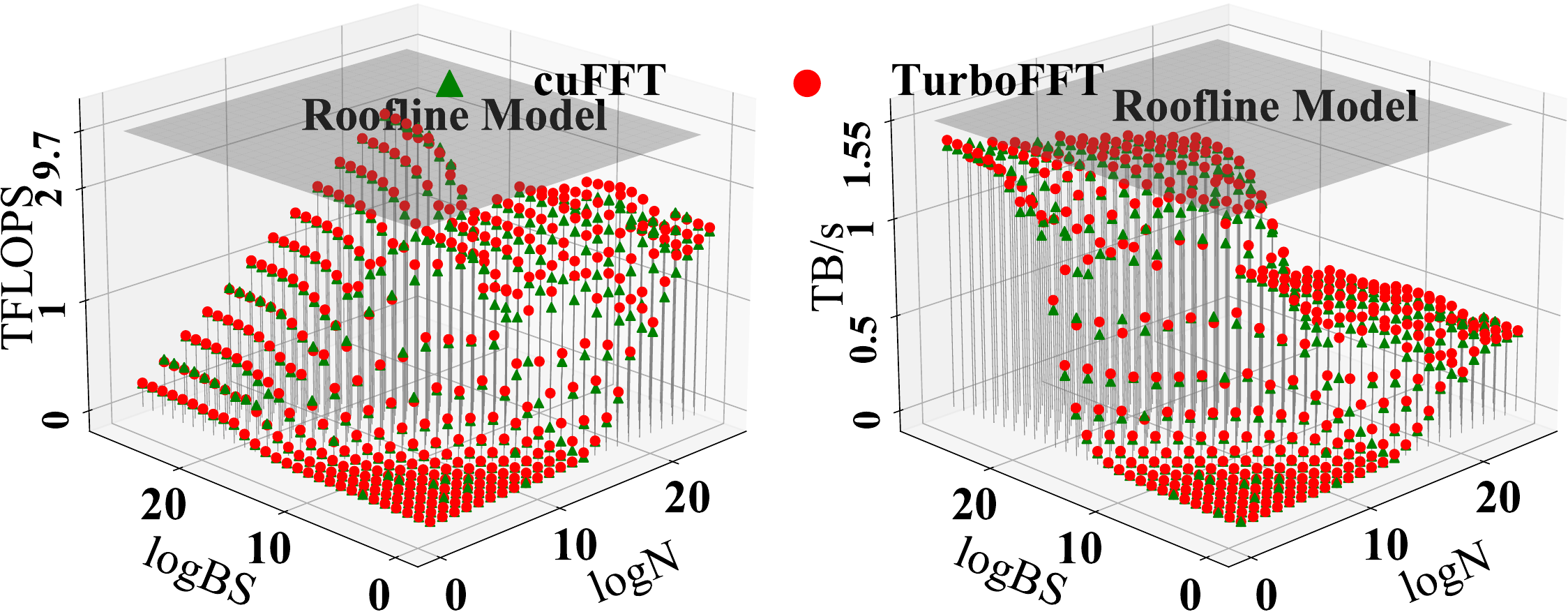}
    \caption{Performance of generated FP64 FFT kernels on A100}
    \label{fig:bench_FP64_a100}
    \vspace{-1mm}
\end{figure}

The code generation strategy offers flexibility to support a wide range of input shapes and datatypes. Figure \ref{fig:bench_FP32_a100} demonstrates the performance of FP32 kernels generated by TurboFFT on A100 GPU. Green point stands for cuFFT, while the red points stand for TurboFFT. The x-y plane demonstrates the input sizes, while the z-axis denotes the performance. We employ TFLOPS and TB/s to measure the compute performance and memory bandwidth. The grey plane stands for the roofline model of the hardware. As Figure \ref{fig:bench_FP32_a100} illustrated, TurboFFT maintains a negligible overhead of $0.58\%$ on average compared to cuFFT. The comparison of FP64 FFT in Figure \ref{fig:bench_FP64_a100} demonstrates TurboFFT has a 7.75\% overhead on average compared to cuFFT.

\subsubsection{Bottleneck Analysis} 
The bottleneck of FFT can be categorized into computation, shared memory, and global memory.
\textit{Computation.} Computation bottleneck, primarily due to slow clock cycles caused by trigonometric functions or double precision operations. Therefore, we pre-calculate and store the required trigonometric function values in global memory. The workload for each thread can be configured in our code generation strategy.

\noindent\textit{Shared memory.} For smaller FFTs, the bottleneck mainly lies in potential shared memory bank conflicts and the FFT size allocated per thread. The issue of shared memory bank conflict primarily originates from the first twiddling process, where each thread needs to access a continuous memory region, an operation that easily causes different threads within the same warp to access the same memory bank. VkFFT uses padding, namely skip one bank per 8 or 16 threads. Although this method can avoid bank conflicts, it wastes a significant amount of shared memory, leading to decrement of threadblocks per SM and performance loss in cases with larger N. Padding can be replaced with swizzling. This is feasible thanks to the unit memory transaction size for a single thread of C2C and Z2Z FFTs being 8 bytes or 16 bytes. We observed that this operation yields a 20\% performance improvement when N is small. For FFT sizes that require more than two kernel launches, it's unnecessary to consider bank conflicts. This is because a single warp's threads can be assigned to different batches. By setting the batch\_id as an offset, threads within the same warp can completely avoid the possibility of bank conflicts.

\noindent\textit{Global memory.} Although for smaller FFT sizes, each thread block only needs to launch a small number of threads, when the batch size increases, each thread block should also increase the number of threads launched correspondingly to enhance throughput. For instance, as shown in the diagram, when $\log N$ ranges from 0 to 5, the throughput of the FFT kernel rapidly rises to over 80\% efficiency with the increase in batch size. Global memory demands high cache efficiency. We found that inefficient access methods on the A100 can lead to up to a 100\% loss in L1 cache hit rate, resulting in nearly 10,000 cycles of stall time for each thread block launch. This is especially noticeable in larger FFT computations, such as when $N=2^{24}$. Due to the need for three launches, the kernel in the final launch requires an additional transposition operation, changing the storage in memory from the original $(N_1, N_2, N_3)$ to $(N_3, N_2, N_1)$. Although allocating thread blocks along the $(N_1, N_2)$ direction maximizes data locality, it leads to writing back along the direction with the largest stride $(N2, N1)$, resulting in a significantly high rate of L1-cache misses and a 30\% overhead.

% Figure \ref{fig:bench_FP32_a100} demonstrates how our TurboFFT kernel is optimized from 3\% to 99\% of cuFFT stepwise. The performance is measured with GFLOPS (bar plot, the left y-axis) and the performance ratio with respect to cuFFT (line chart, the right y-axis). For the most basic version, TurboFFT-v0, each thread handles a radix-2 FFT, and a $log_2(N)$ times of kernel launches are required. Without using any optimizations, the TurboFFT-v0 obtains a performance of 49 GFLOPS. Next in TurboFFT-v1, we enlarge the thread-level radix, namely the thread-level FFT size, from $2$ to $16$. The performance improved from 49 GFLOPS to 110 GFLOPS. And then in TurboFFT-v2, FFT with a size larger the 8192 is decomposed into 2 or 3  batched-FFT to reduce the number of kernel launches from $log_2(N)$ to 2 or 3. The decomposition enables each threadblock to perform several batches of the small FFT independently using the shared memory. The performance is improved to 334 GFLOPS. After that, the final high-performance version TurboFFT exploits the shared memory again to reorder and coalesce the read/write of the FFT input/output from/to global memory. The performance of TurboFFT achieves 565 GFLOPS. By employing the above thread-level radix upscaling, threadblock-level FFT decompositions, and reordering using shared memory, we finally improved the performance ratio from 5\% to 99\% with respect to cuFFT. 

\subsection{Benchmarking TurboFFT with Fault Tolerance}
Our optimizations on TurboFFT from scratch, including tiling, thread-level computation balancing, twiddling factor precomputing, and template-based flexibility in kernel parameter selection enable further development of lightweight fault-tolerance for TurboFFT.
\begin{figure}[tp]
\vspace{-2mm}
    \centering
\includegraphics[width=1\linewidth]{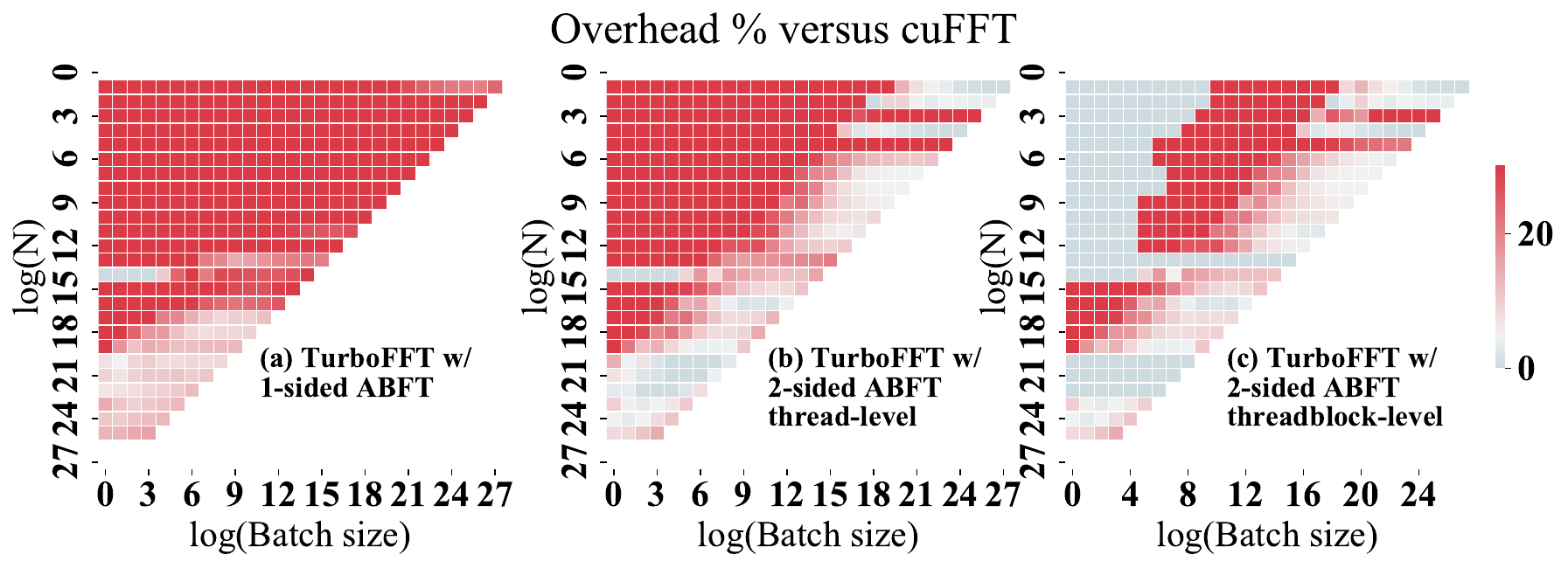} 
    \caption{Different schemes of 2-sided ABFT for FP32 FFT on A100.}
    \label{fig:twoside_abft_stepwise_optimization_A100_FP32}
    \vspace{-1mm}
\end{figure}

Figure \ref{fig:twoside_abft_stepwise_optimization_A100_FP32} offers an evaluative comparison across three error detection strategies. From left to right in this figure, we can clearly observe a reduction in overhead. Specifically, Figure \ref{fig:twoside_abft_stepwise_optimization_A100_FP32} (a) showcases a one-sided FFT with an ABFT scheme, noted for its 29\% overhead when benchmarked against the cuFFT. Moving on to Figure \ref{fig:twoside_abft_stepwise_optimization_A100_FP32} (b), it introduces a two-sided ABFT at the thread level, demonstrating a reduced overhead of 13.38\%. Finally, Figure \ref{fig:twoside_abft_stepwise_optimization_A100_FP32} (c) illustrates the efficiency of a two-sided ABFT at the threadblock level, which further lowers the overhead to 8.9\%. The heatmaps within each figure utilize a color spectrum where red indicates a higher overhead and blue a lower one, clearly visualizing the efficiency gains of each method as the color gradation shifts from red to blue. This improvement is attributed to the implementation of a batch detection mechanism, which obviates the need for individual checksums on each signal, thereby circumventing threadblock-level communication and synchronization. Conversely, in the two-sided scheme, we apply a linear combination to a batch of signals, confining the bulk of the computations within individual threads. It's only at the final step that a threadblock-level reduction is performed, significantly minimizing overhead.
\begin{figure}[hp]
\vspace{-3mm}
    \centering
\includegraphics[width=1\linewidth]{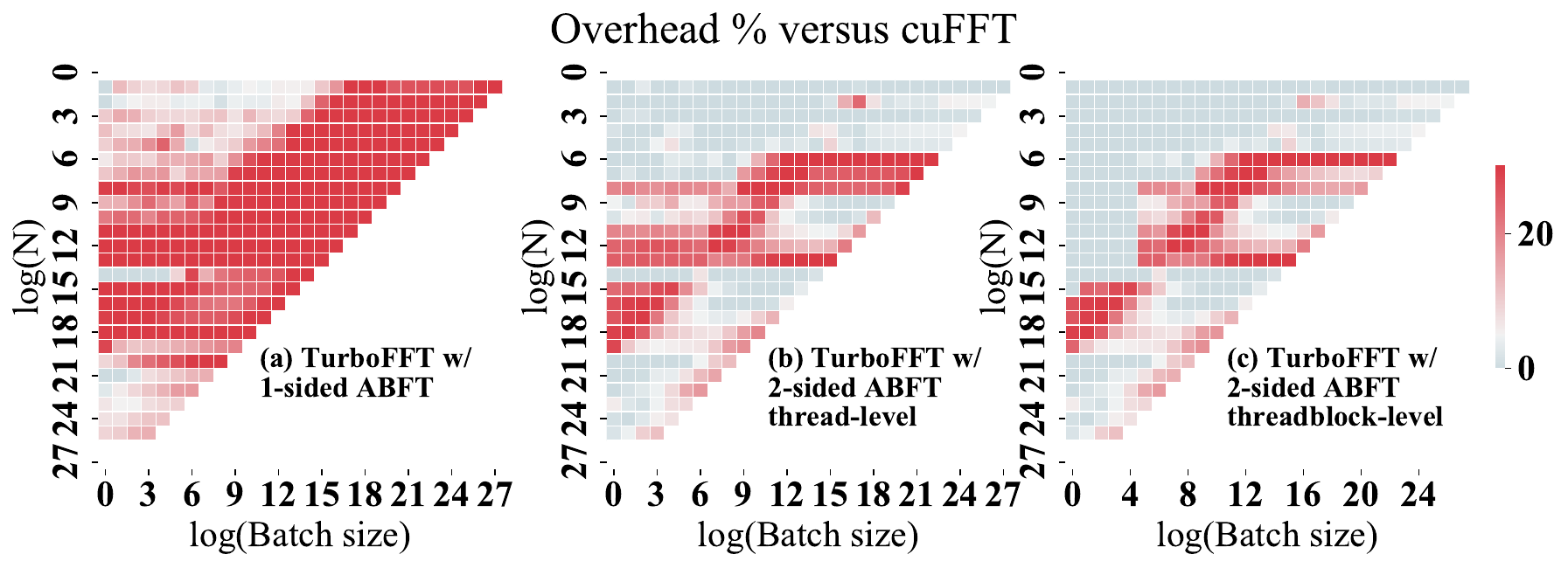} 
    \caption{Different schemes of 2-sided ABFT for FP64 FFT on A100.}
    \label{fig:twoside_abft_stepwise_optimization_A100_FP64}
    \vspace{-2mm}
\end{figure}
Figure \ref{fig:twoside_abft_stepwise_optimization_A100_FP64} replicates the experiment for double precision values. Similarly, we observe a sequential decrease in the area covered by red from left to right. In Figure \ref{fig:twoside_abft_stepwise_optimization_A100_FP64}, parts (a), (b), and (c) showcase checksum overheads of 27.40\%, 10.12\%, and 7.87\%, respectively. This pattern indicates a consistent reduction in overhead as we move through the different methods of error detection and correction.

\begin{figure}[tph]
\vspace{-3mm}
    \centering
    \includegraphics[width=\linewidth]{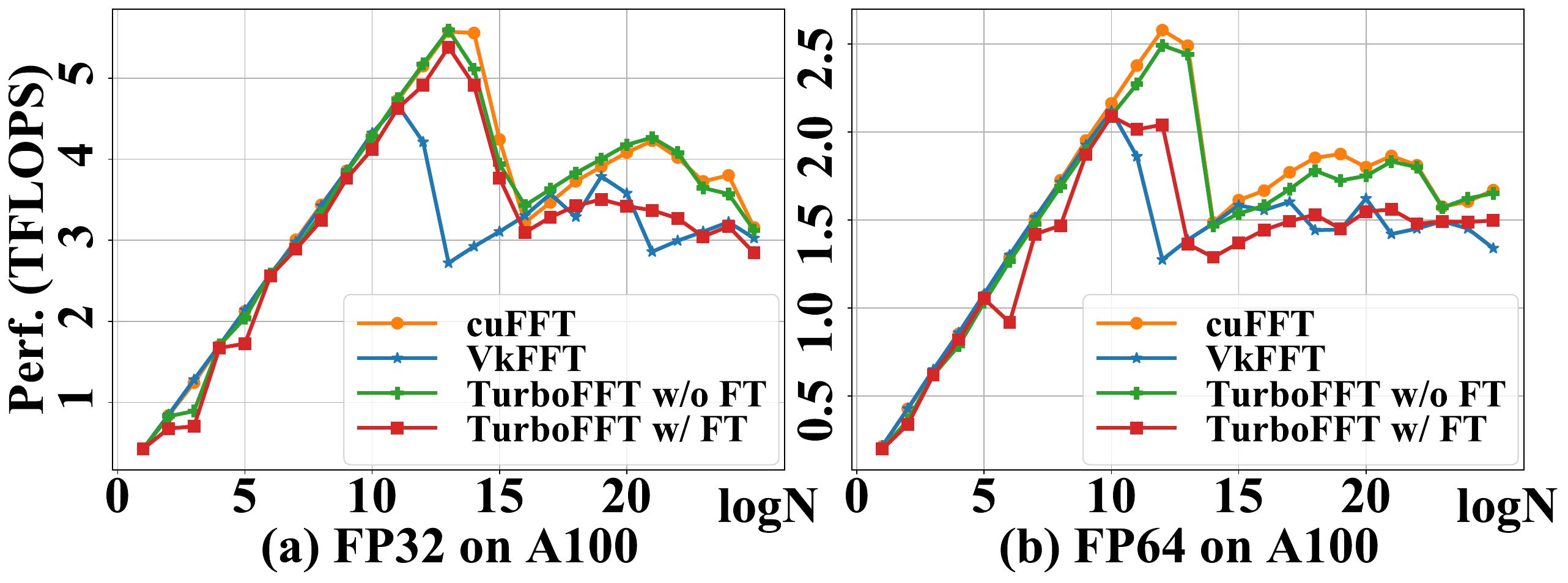}
    \caption{Comparison of TurboFFT performance with and without fault tolerance on an A100 GPU. cuFFT and VkFFT are included. The number of elements is maintained as $2^{28}$.}
    \label{fig:a100_ft}
    \vspace{-1mm}
\end{figure}
Figure \ref{fig:a100_ft} provides a comparison of the performance of TurboFFT in single precision, both without fault tolerance and with fault tolerance, and also includes the performance metrics for cuFFT and VkFFT. The implementation of two-sided checksums results in only an average overhead of 8\% for FP32 and 10\% for FP64 compared to TurboFFT without fault tolerance, which further substantiates the efficiency of this scheme. In addition to the previous findings, TurboFFT equipped with two-sided checksums achieved an overhead of just 10\% relative to cuFFT, which is on par with the overhead observed for VkFFT compared to cuFFT. This indicates not only the high performance of TurboFFT but also the efficiency of the two-sided checksum approach.

\subsection{Benchmarking TurboFFT under Error Injection}

% \begin{subfigure}[ht]
% % \vspace{-4mm}
%     \centering
% \includegraphics[scale=0.22]{figs/fig5_error_injection_T4.pdf} 
%     \caption{Error Injections, T4.}
%     \label{fig:err_T4}
%     \vspace{-4mm}
% \end{subfigure}

\subsubsection{Error Injection Analysis}

\begin{figure}[tp]
\vspace{-4mm}
    \centering
\includegraphics[scale=0.24]{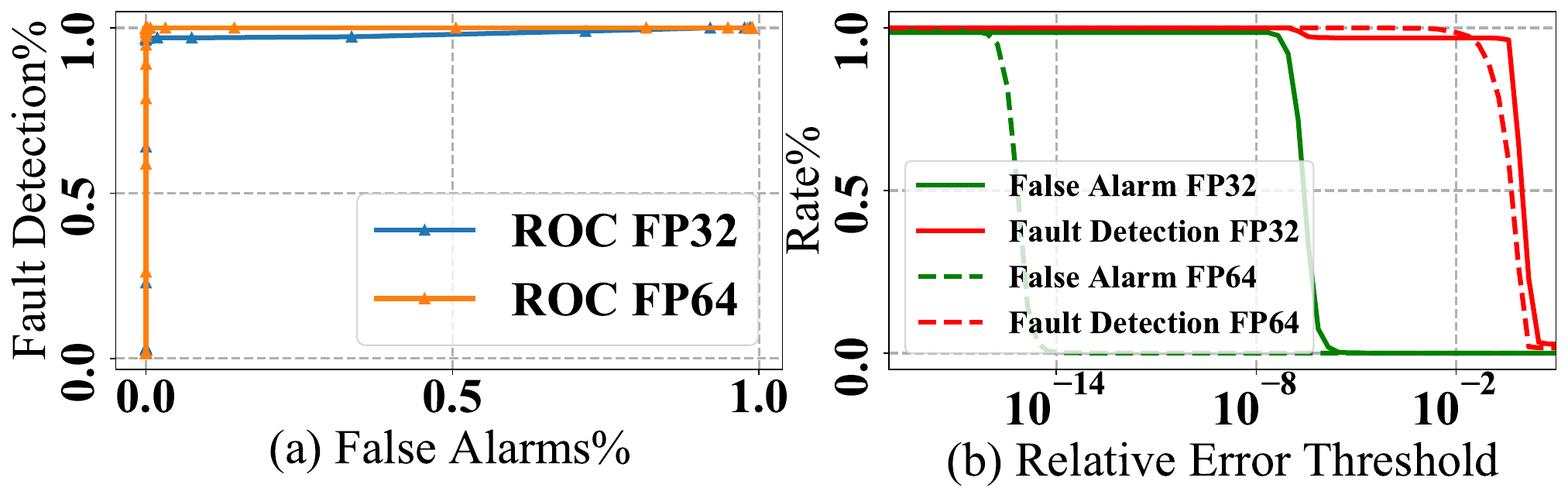} 
    \caption{Error Analysis}
    \label{fig:error_analysis}
    \vspace{-4mm}
\end{figure}
Figure \ref{fig:error_analysis} illustrates that, by selecting an appropriate fault detection threshold $\delta$, the proposed detection scheme is capable of identifying injected faults with a high degree of reliability and a negligible false alarm rate. 2000 random test signals are generated with normal distribution. Faults are injected in half of these runs (1000 of 2000) by first choosing a signal to affect, and then flipping exactly one bit of its 32-bit representation for float-precision and 64-bit representation for double-precision. A checksum test with threshold $\delta$ is used to attempt to identify the affected computations. Characteristics of the proposed fault detection scheme are demonstrated using the standard receiver operating characteristic (ROC) curve in Figure \ref{fig:error_analysis} (a). For a given fault threshold $\delta$, a proportion of False Alarms (numerical errors greater than $\delta$ tagged as data faults) and Detections (injected data faults correctly identified) will be observed. The ROC curve parametrically maps these two proportions as the tolerance level adjusts. Figure \ref{fig:error_analysis} (b) presents the detection rate and false alarm rate versus the fault detection threshold.

\subsubsection{Error Injection Evaluation}

% We focus on compute errors rather than memory errors in the paper. Errors are inserted in the register of FFT signals by adding a numerical offset to emulate register bit flipping. For each thread block, errors are evenly distributed throughout the computation, with the interval between each insertion error being K/20, where K is the FFT batch size. Each error is inserted in a random thread. The injected error will lead to a mismatch in the checksum detection phase, such that the erroneous element and error magnitude can be computed according to the two-sided ABFT scheme. The threadblock which corrects the error then stores back the result. After detecting no errors in the result, we validate the correctness by comparing them with cuFFT to ensure the sanity of our protection schemes.
\begin{figure}[th]
    \vspace{-3mm}
    \centering
    \includegraphics[width=\linewidth]{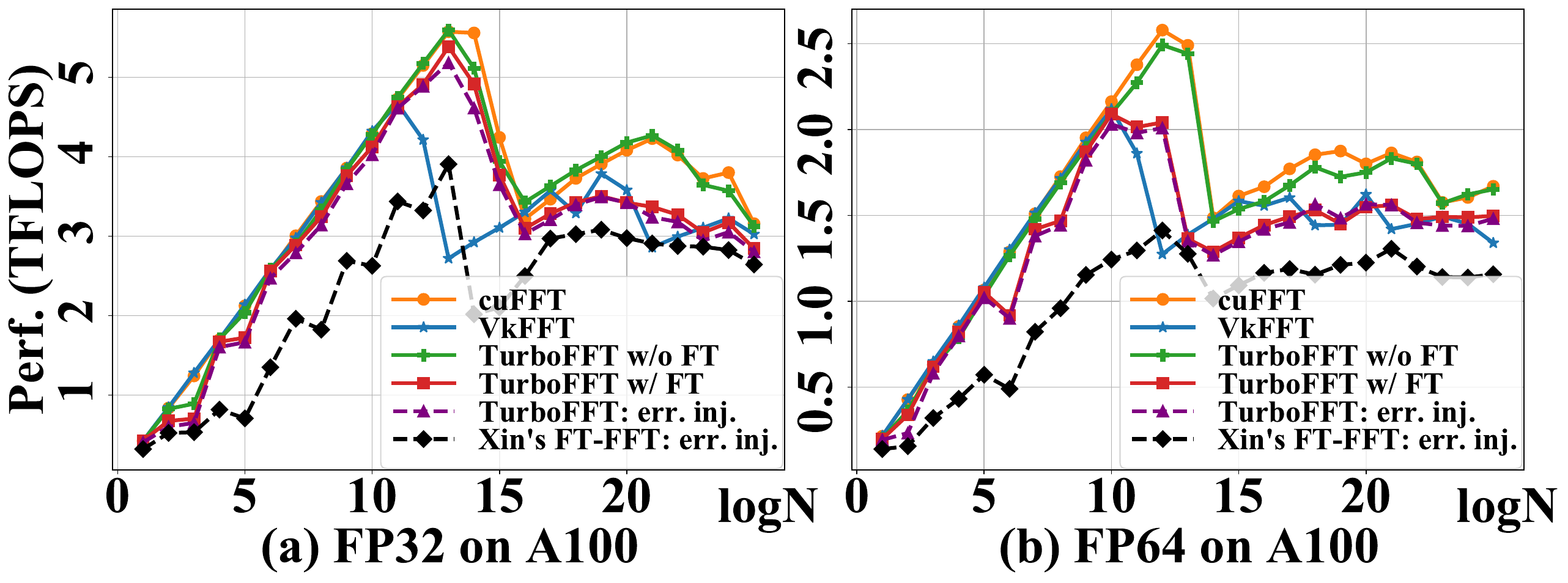}
    \caption{Error injection experiments on A100 GPU}
    \label{fig:err_inj_A100_FP32}
    \vspace{-3mm}
\end{figure}

Figures \ref{fig:err_inj_A100_FP32} extends the analysis to include the performance of TurboFFT under error injection scenarios, and introduces Xin's method for comparison. The figure reveals that TurboFFT, when subjected to error injection, incurs a negligible overhead of 3\% for FP32 and 2\% for FP64 compared to scenarios without error injection. Using cuFFT as a baseline, the overhead for TurboFFT with error injection stands at 13\%, whereas Xin's method exhibits a significantly higher overhead of 35\% relative to cuFFT. 

\subsection{Performance Evaluation on T4}
\begin{figure}[h]
\vspace{-3mm}
    \centering
    \includegraphics[width=1\linewidth]{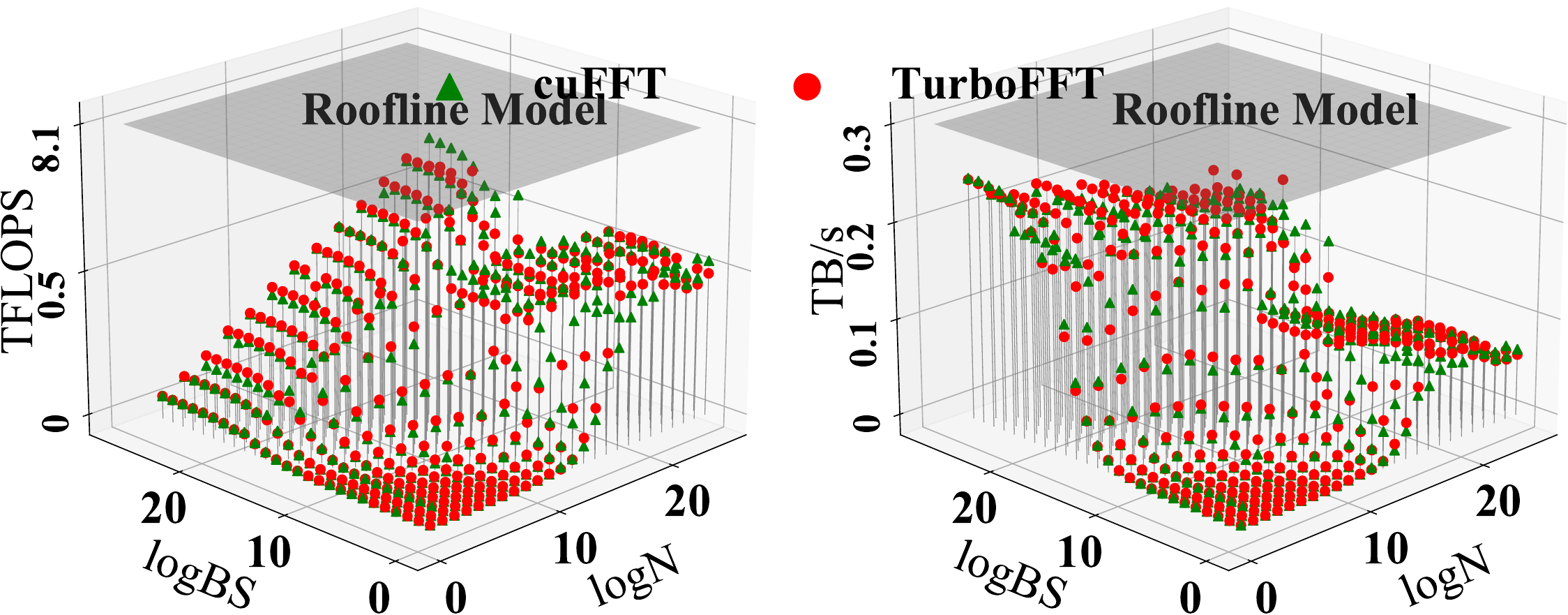}
    \caption{TurboFFT vs cuFFT on T4, FP32}
    \label{fig:bench_FP32_t4}
    \vspace{-3mm}
\end{figure}

\begin{figure}[t]
% \vspace{-5mm}
    \centering
    \includegraphics[width=1\linewidth]{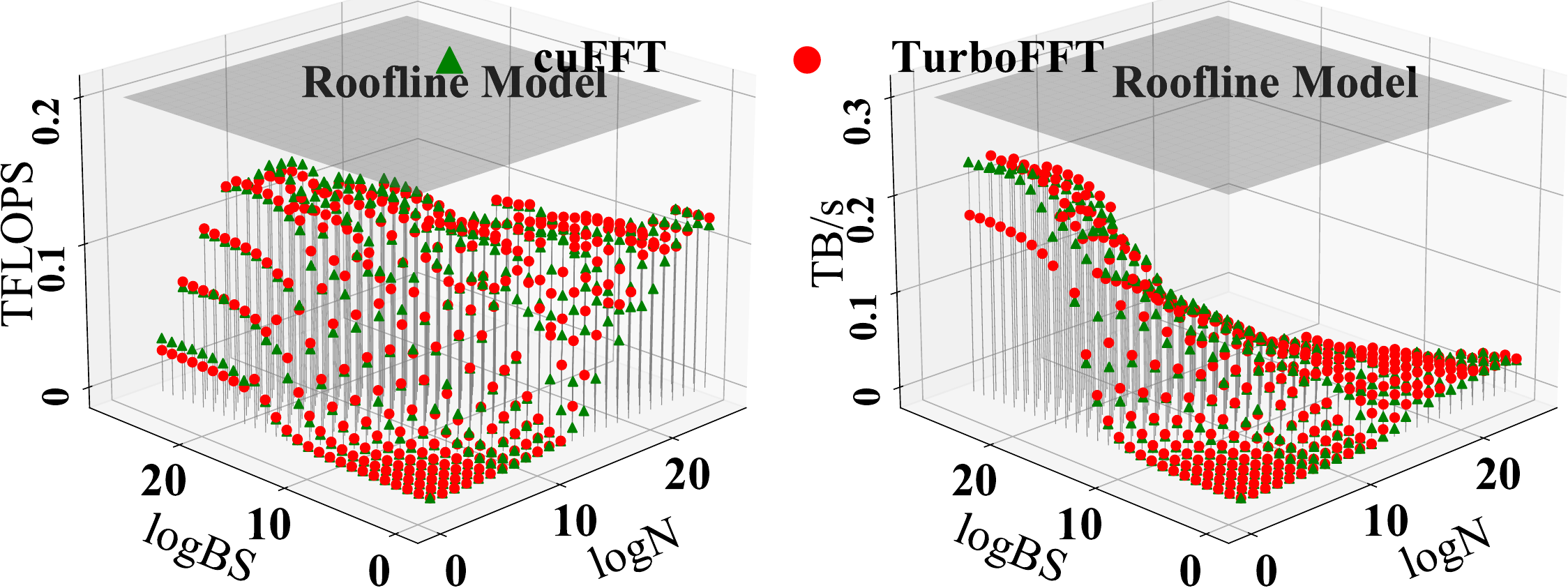}
    \caption{TurboFFT vs cuFFT on T4, FP64}
    \label{fig:bench_FP64_t4}
    \vspace{-1mm}
\end{figure}

In addition to conducting evaluations on the NVIDIA A100 GPU, the effectiveness of the optimizations has also been tested on an NVIDIA Tesla T4 GPU. Figure \ref{fig:bench_FP32_a100} demonstrates the performance of FP32 kernels generated by TurboFFT on A100 GPU. As Figure \ref{fig:bench_FP32_t4} illustrated, TurboFFT maintains a negligible overhead of 3.77\% on average compared to cuFFT. The comparison of FP64 FFT in Figure \ref{fig:bench_FP64_t4} demonstrates TurboFFT has a 7.63\% overhead on average compared to cuFFT. Furthermore, Figure \ref{fig:bench_FP64_t4} showcased the poor support of double precision on T4 given it is a GPU card specially designed for deep learning. Even with large batch sizes and FFT sizes, both memory throughput and compute performance do not exceed 300 GB/s and 200 GFLOPS, respectively. Consequently, given that FP64 performance has already been assessed on the A100, our focus on the T4 has been directed towards testing FP32 performance.
% In addition to cuFFT, we further evaluate the effectiveness of our optimizations vs VkFFT and our fault tolerance scheme vs Xin's FT-FFT \cite{liang2017correcting} on A100. 
% Although Xin's FT-FFT is a CPU-based ABFT FFT scheme, we implement the algorithm inside TurboFFT first. Figure \ref{fig:benchmark_A100} illustrates how our TurboFFT outperforms Xin's FT-FFT in fault tolerance by approximately 47\% across varied fault protection schemes. Xin's FT-FFT, although employing one-sided ABFT, is limited by excessive memory operations. This excessive use of memory bandwidth leads to a decrease in performance. In contrast, the two-sided ABFT in TurboFFT uses warp shuffle to minimize memory overhead as much as possible, resulting in an overhead of 12\%, compared to the 53\% in Xin's FT-FFT.

\begin{figure}[htp]
\vspace{-2mm}
    \centering
\includegraphics[width=1\linewidth]{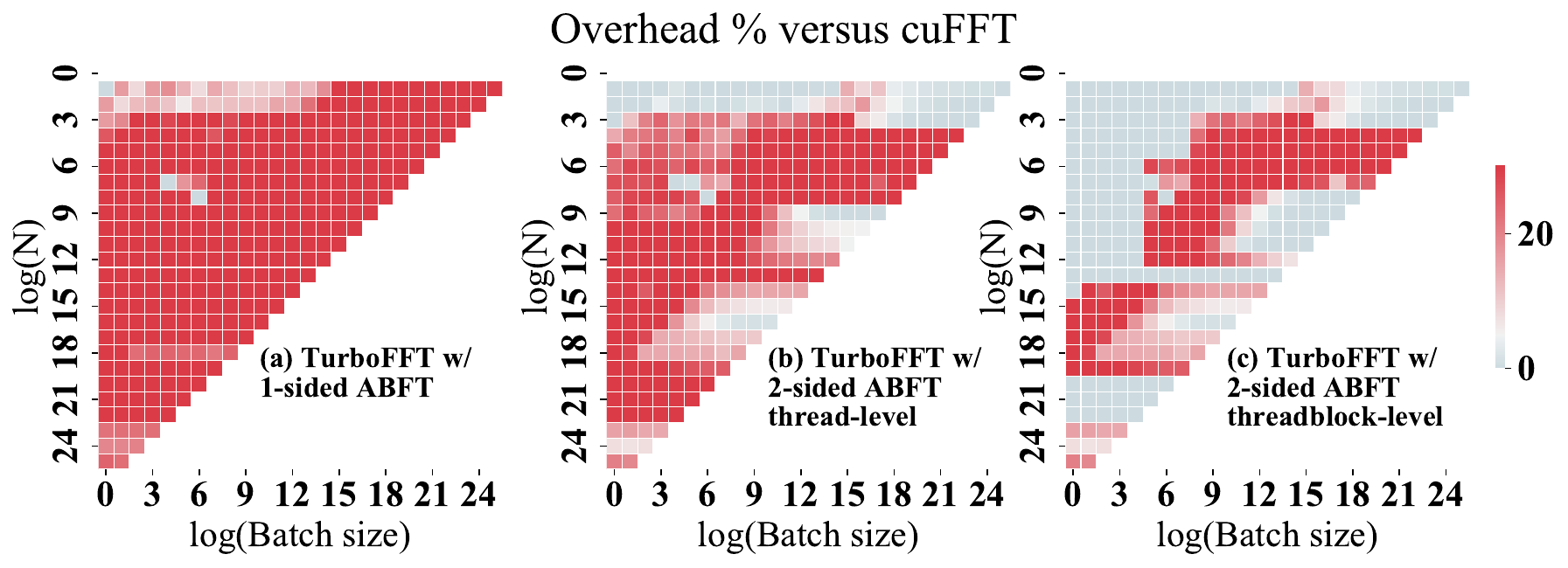} 
    \caption{Stepwise Optimization of 2-sided ABFT for FP32 FFT on T4.}
    \label{fig:twoside_abft_stepwise_optimization_T4_FP32}
    \vspace{-2mm}
\end{figure}
Figure \ref{fig:twoside_abft_stepwise_optimization_T4_FP32} offers an evaluative comparison across one-sided checksum, thread-level two-sided checksum, and threadblock-level two-sided checksum. From left to right in this figure, we can clearly observe a reduction in overhead. Specifically, Figure \ref{fig:twoside_abft_stepwise_optimization_T4_FP32} (a) showcases a one-sided FFT with an ABFT scheme, noted for its 45.68\% overhead when benchmarked against the cuFFT. Moving on to Figure \ref{fig:twoside_abft_stepwise_optimization_A100_FP32} (b), it introduces a two-sided ABFT at the thread level, demonstrating a reduced overhead of 25.94\%. Finally, Figure \ref{fig:twoside_abft_stepwise_optimization_A100_FP32} (c) illustrates the efficiency of a two-sided ABFT at the threadblock level, which further lowers the overhead to 15.01\%. 

\begin{figure}[h]
    \centering
    \includegraphics[width=0.9\linewidth]{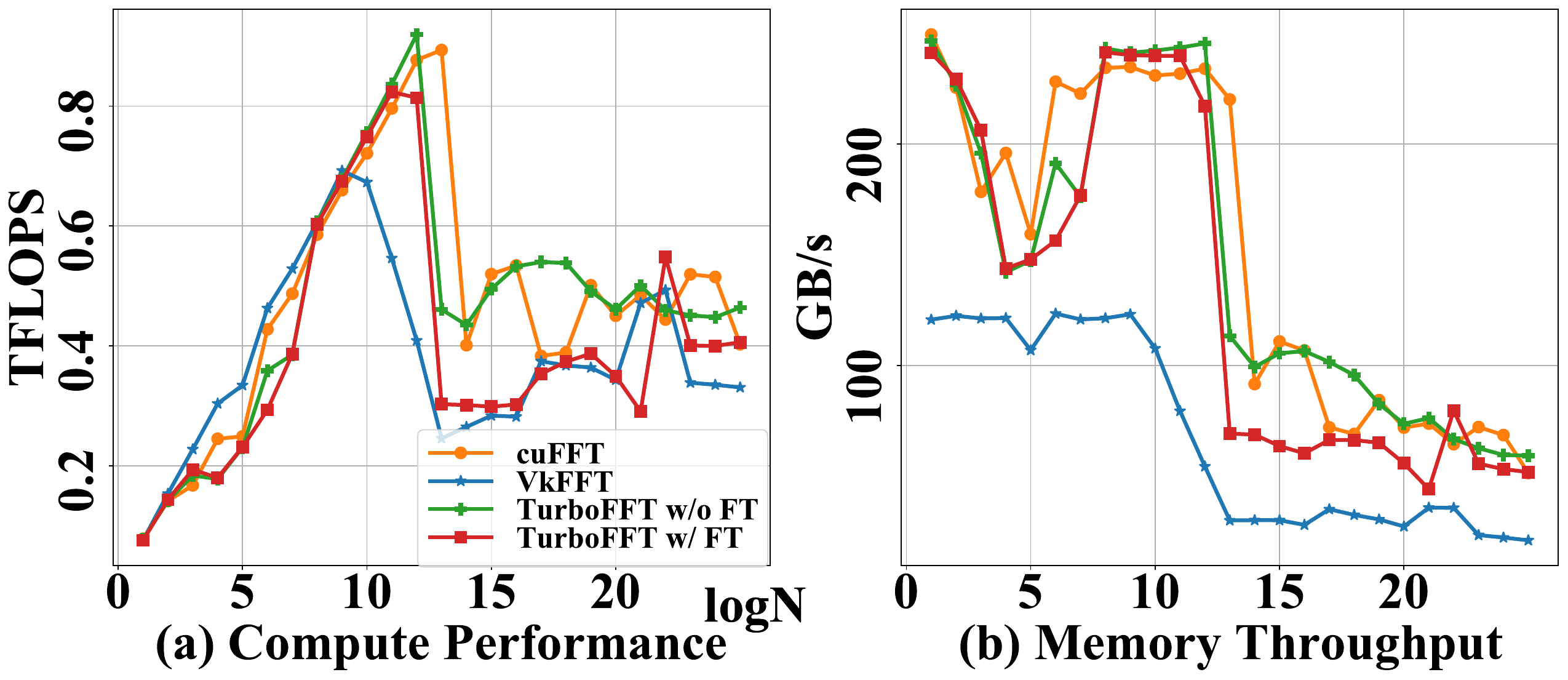}
    \caption{Comparison of TurboFFT FP32  performance with and without fault tolerance on a T4 GPU. cuFFT and VkFFT are included. The number of elements is maintained as $2^{28}$.}
    \label{fig:t4_ft}
    \vspace{-1mm}
\end{figure}
Figure \ref{fig:t4_ft} provides a comparison of the performance of TurboFFT in single precision, both without fault tolerance and with fault tolerance, and also includes the performance metrics for cuFFT and VkFFT. The TurboFFT without fault tolerance maintains a performance close to cuFFT, outperforming VkFFT, which has a 12\% overhead compared to cuFFT. The implementation of two-sided checksums results in an average overhead of 14\% compared to TurboFFT without fault tolerance.
\begin{figure}[htp]
    \vspace{-1mm}
    \centering
    \includegraphics[width=0.9\linewidth]{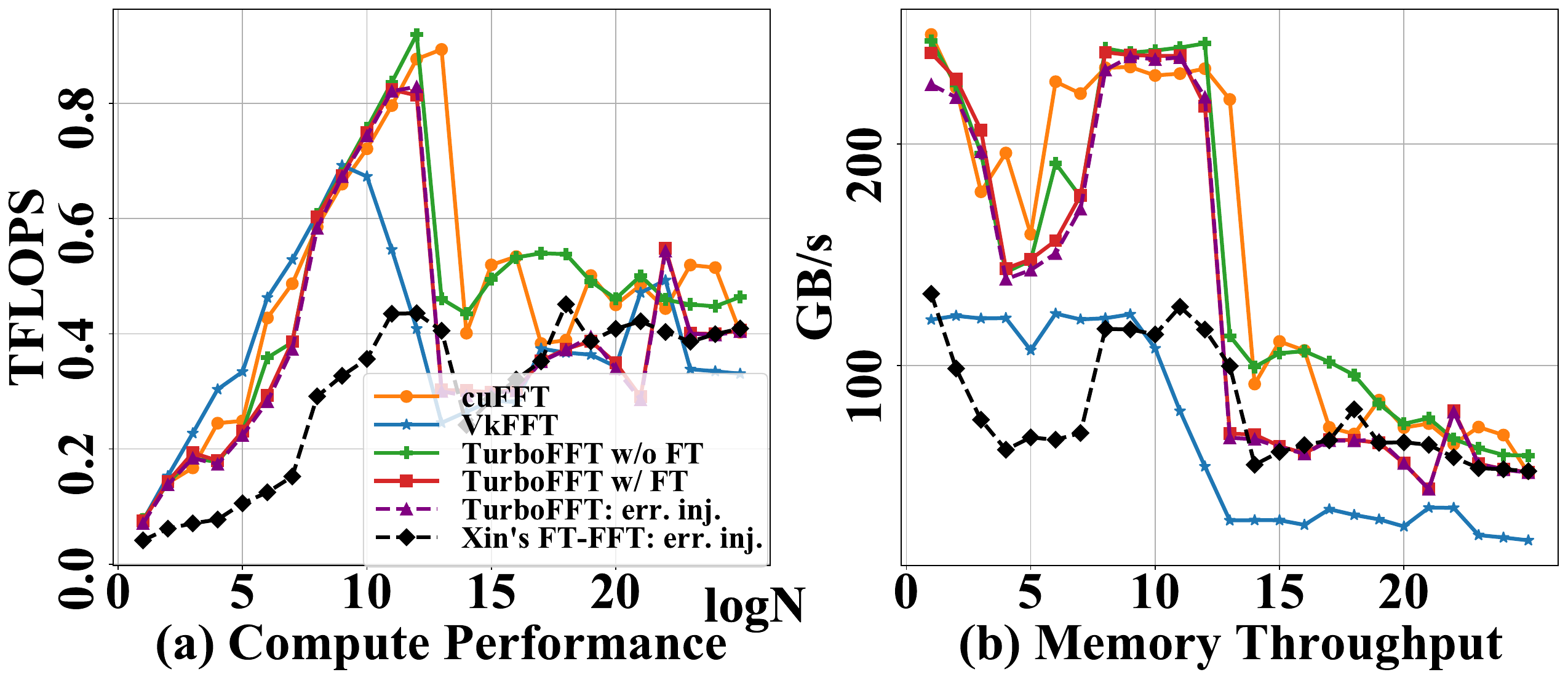}
    \caption{Error injection experiments on T4 GPU}
    \label{fig:err_inj_T4_FP32}
    \vspace{-1mm}
\end{figure}

Figures \ref{fig:err_inj_T4_FP32} details the comparison of TurboFFT under error injection on T4, and Xin's FT-FFT is included. As shown in Figures \ref{fig:err_inj_T4_FP32}, TurboFFT under error injection incurs a negligible overhead of 3\% for FP32 and 2\% for FP64 compared to scenarios without error injection. Using cuFFT as a baseline, the overhead for TurboFFT with error injection stands at 16\%, achieving more than 1 times speed up compared to the 38\% overhead introduced by the one-sided fault tolerance scheme used in Xin's FT-FFT. 

\section{Conclusion}
\label{sec:conclusion}
 In this paper, we introduce \textit{TurboFFT}, a lightweight FFT implementation, which not only provides architecture-aware fault tolerance optimizations but also overcomes error propagation and enables on-the-fly fault correction, eliminating the need for routine recomputation. We address the error propagation issue by proposing a novel \textit{two-sided} checksum scheme, in conjunction with a series of architecture-aware optimizations specifically designed for GPU. The proposed \textit{two-sided} checksum involves calculating checksums along individual signals for detecting errors and across different signals for correcting errors. A code generation strategy is employed to support and boost the performance for a variety of input sizes, data types. Experimental results on an NVIDIA A100 server GPU and Tesla Turing T4 GPUs show that TurboFFT holds a competitive performance compared to the state-of-the-art closed-sourced library, cuFFT. The fault tolerance scheme in TurboFFT maintains a low overhead (7\% to 15\%), even under hundreds of error injections per minute for both single and double precision.
% \section{Acknowledgement}

% This work was supported by the U.S. Department of Energy, Office of Science, Office of Advanced Scientific Computing Research, Scientific Discovery through the Advanced Computing (SciDAC) program under Award Number DE-SC0022209. We thank the anonymous reviewers for their insightful comments.

\newpage
\renewcommand*{\bibfont}{\footnotesize}
\printbibliography[]

%\newpage
% \setcounter{tocdepth}{4}
%\renewcommand{\contentsname}{ToC, not part of the submission}
%{\footnotesize\sffamily\tableofcontents}

\end{document}